\documentclass[aps,prb,superscriptaddress,showpacs,longbibliography,reprint]{revtex4-2}

\usepackage[colorlinks=true,linkcolor=red,urlcolor=blue,citecolor=blue,pdfusetitle]{hyperref}
\usepackage{color}
\usepackage[utf8]{inputenc}
\usepackage[usenames,dvipsnames]{xcolor}
\usepackage{amsmath}
\usepackage{physics}
\usepackage{amssymb}
\usepackage{graphicx}
\usepackage{tikz}
\usepackage{pgfplots}
\usepackage{pgfplotstable}
\graphicspath{{graphics/}}
\usepackage{epsfig}
\usepackage{dcolumn}
\usepackage{bm}
\usepackage{mathrsfs}
\usepackage{multirow}
\usepackage[all]{xy}
\usepackage{pbox}
\usepackage{lipsum}
\usepackage{verbatim}
\usepackage{braket}
\usepackage{dsfont}
\usepackage{array}
\usepackage{makecell}
\usepackage{tabularx}
\usepackage{isotope}
\usepackage{xr}
\usepackage{float}
\usepackage[english]{babel}
\usepackage{csquotes}
\usepackage{listings}
\usepackage{comment}
\usepackage{setspace}
\usepackage{mathtools}
\usepackage{float}

\pgfplotsset{compat=1.18}

\usepackage[caption=false]{subfig}
\usepackage{ragged2e} 
\DeclareCaptionJustification{justified}{\justifying}

\definecolor{deepjunglegreen}{rgb}{0.0, 0.29, 0.29}

\newcommand{\tubingen}{Institut f\"{u}r Theoretische Physik,  Universit\"{a}t T\"{u}bingen, Auf der Morgenstelle 14, 72076 T\"{u}bingen, Germany}

\newcommand{\nottingham}{School of Physics and Astronomy and Centre for the Mathematics and Theoretical Physics of Quantum Non-Equilibrium Systems, The University of Nottingham, Nottingham, NG7 2RD, United Kingdom}
\newcommand{\paris}{Laboratoire de Physique Théorique et Modèles Statistiques, Université Paris-Saclay, CNRS, 91405 Orsay, France}

\begin{document}
\raggedbottom

\title{Large-scale universality in quantum reaction-diffusion from Keldysh field theory}

\author{Federico Gerbino}
\email{federico.gerbino@universite-paris-saclay.fr}
\affiliation{\paris}

\author{Igor Lesanovsky}
\affiliation{\tubingen}
\affiliation{\nottingham}

\author{Gabriele Perfetto}
\affiliation{\tubingen}

\begin{abstract}
We consider the quantum reaction-diffusion dynamics in $d$ spatial dimensions of a Fermi gas subject to binary annihilation reactions $A+A \to \emptyset$. These systems display collective nonequilibrium long-time behavior, which is signalled by an algebraic decay of the particle density. Building on the Keldysh formalism, we devise a field theoretical approach for the reaction-limited regime, where annihilation reactions are scarce. Combining a perturbative 
expansion of the dissipative interaction with Euler-hydrodynamic scaling limit, we derive a description in terms of a large-scale universal kinetic equation.
Our approach shows how the time-dependent generalized Gibbs ensemble assumption, which is often employed for treating low-dimensional nonequilibrium dissipative systems, emerges from systematic diagrammatics. It also allows us to exactly compute---for arbitrary spatial dimension---the decay exponent of the particle density. The latter is based on the large-scale description of the quantum dynamics and it differs from the mean-field prediction even in dimension larger than one. We moreover consider spatially inhomogeneous setups involving an external potential. In confined systems the density decay is accelerated towards the mean-field algebraic behavior, while for deconfined scenarios the power-law decay is replaced by a slower nonalgebraic decay. 
\end{abstract}

\maketitle

\textit{Introduction.}  
Reaction-diffusion (RD) systems \cite{henkel2008non, hinrichsen2000non, vladimir1997nonequilibrium}, where particles diffuse and react upon meeting, 
are ideal systems for the investigation of dynamical universal behavior. 
For example, for binary annihilation reactions $A+A\to \emptyset$, the late time decay of the particle density takes a universal power-law form. In the ``diffusion-limited'' regime  \cite{OVCHINNIKOV78, kang84scaling, kang84fluct, kang85,spounge88,torney1983diffusion,privman94,toussaint1983particle,racz1985}, where diffusion is weak, the origin of this dynamical behavior are spatial density fluctuations. Here, mean-field approaches cannot be applied and field-theoretical and renormalization group analyses \cite{doi1976, peliti1985,peliti1986renormalisation, Lee1994, cardy1996, mattis1998,grassberger1980fock,tauber2002dynamic, tauber2005, tauber2014critical} correctly predict the observed power-law. The mean-field approximation is, however, valid in more than one dimension and/or in the ``reaction-limited'' regime of fast hopping mixing \cite{hinrichsen2000non,vladimir1997nonequilibrium,kang84fluct,fastdiffusion1992,Urbano2023}. 

\begin{figure*}
    \centering
    \includegraphics[width=\textwidth]{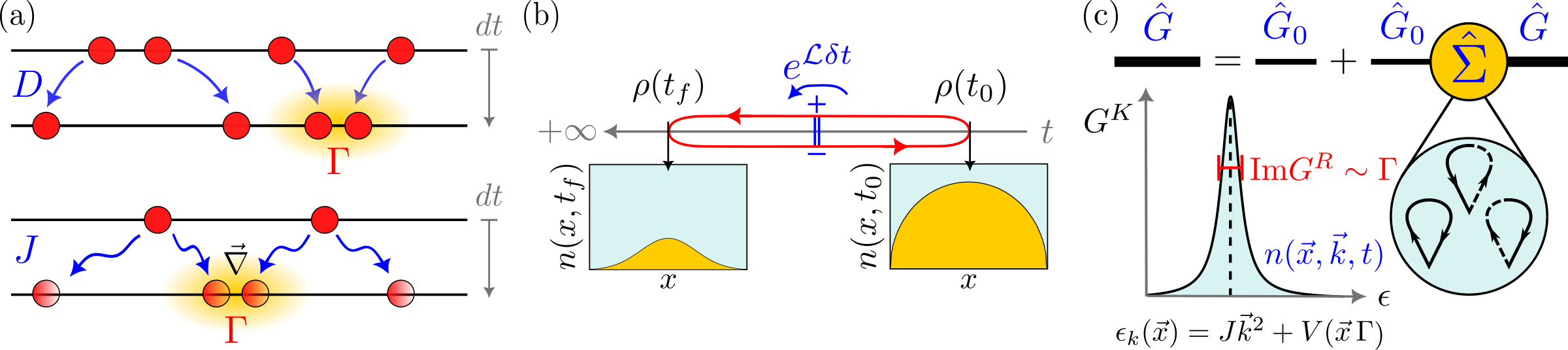}
    \caption{Quantum RD dynamics via Keldysh field theory. (a) Comparison of classical and quantum RD dynamics: classical incoherent diffusion ($D$, top, blue solid lines) is replaced by quantum coherent ballistic motion ($J$, bottom, blue wiggly lines), while in both cases annihilation, $A+A\to \emptyset$, is irreversible. 
    (b) Time evolution of the density matrix $\rho(t)$ along the closed-time contour associated to the Lindblad map $\mathcal{L}$. Two time branches, forward ($+$) and backward ($-$), are required. The particle number decreases during time evolution, as can be seen by comparing the initial $n(x,t_0)$ and final $n(x,t_f)$ density profiles.
    (c) Annihilation interaction vertices define the self-energy Keldysh matrix $\hat{\Sigma}$, which dresses the Green's functions $\hat{G}$ with respect to their bare values $\hat{G}_0$. In the reaction-limited regime $\hbar n(x,t)\Gamma/J\ll 1$, the quasiparticle dispersion relation $\epsilon_k$ is not modified, while quasiparticles acquire a large-finite lifetime $\sim \Gamma^{-1}$, given by the energy width $\epsilon$ of $G^K(\vec{x},t,\vec{k},\epsilon)$ (sketched in red). This lifetime is given by tadpole Feynman diagrams (sketched in the light blue inset). In the Euler-scaling limit, these diagrams reproduce the TGGE predictions for $n(\vec{x},\vec{k},t)$ (spectral integral in $\epsilon$ of $G^K(\vec{x},t,\vec{k},\epsilon)$ depicted in light blue).
    }
    \label{fig:diffusion}
\end{figure*}

In quantum many-body systems large-scale properties are even harder to uncover than in the classical realm, already in the one-dimensional case, since they entail the simulations of large sizes and long times \cite{van_horssen2015,gillman2019,gillman20,jo2021,carollo2019,carollo2022nonequilibrium,carollo2022quantum}. In this regard, quantum RD systems have moved into the focus of attention. They follow simple dynamical rules \cite{lossth7,lossth8,lossth1,Rosso2022,lossth3,lossth4,lossth5,lossth6,perfetto22,Nosov2023,perfetto23,lossth9,lossth10,lossth11}, which connect to cold-atomic experiments \cite{lossexp0,lossexp1, lossexp2, lossexp3, lossexp4, lossexp5, lossexp6, lossexp7,lossexpF1,lossexpF2,lossexpF3}, and they allow for novel forms of particle-density decay beyond mean-field. This has been shown in Refs.~\cite{lossth1,Rosso2022,lossth3,lossth6,perfetto22,perfetto23,lossth11} for one-dimensional systems in the reaction-limited regime. In this limit, analytical predictions can be obtained under the assumption that the systems relaxes to a time-dependent generalized Gibbs ensemble (TGGE) \cite{tGGE1,tGGE2,tGGE3,tGGE4}. The connection between the TGGE assumption and diagrammatics techniques, and the study of universal decay in generic spatial dimensions $d$, requires, however, the development of a field theory.

In this Letter, we accomplish this by exploiting the Keldysh path integral representation of the quantum master equation \cite{schwinger1961brownian, keldysh1965, kamenev2011, altland2010,sieberer2016,tonielli2016,thompson2023field,sieberer2023universality}. We investigate as a paradigmatic example the Fermi gas in $d$ spatial dimensions subject to binary annihilation reactions $A+A\to\emptyset$. From the Keldysh field theory, we perform a diagrammatic expansion of the dissipative interaction vertices. In the Euler-scaling limit of hydrodynamics \cite{spohn2012large,Dojon20,essler2022short,lossth5,de2022correlation}, when space-time derivatives are kept at leading order, this expansion acquires the universal form of a kinetic Boltzmann equation. In $d=1$, this analysis provides results equivalent to that of the TGGE ansatz for the reaction-limited regime and therefore provides a connection between the much-employed TGGE relaxation assumption, hydrodynamic scaling limits, and diagrammatic expansions in dissipative systems.
Crucially, from the field-theory description, we exactly compute the density decay exponent in arbitrary dimensions, which is found to deviate from the mean-field prediction even in $d>1$. This result is in contrast with the classical case and it is rooted into the large-scale universal description of the underlying quantum dynamics. We also consider the case of inhomogeneous systems where we study quenches of a trapping potential confining 
the fermions. For a quench from a double to single well potential, we find an acceleration of the particle decay, which diverts the decay exponent towards the mean-field one. For a trap-release quench, we, instead, find a qualitatively different scenario: 
the algebraic decay first slows down on an intermediate time window, and then it gets replaced at long times by a slower non-algebraic decay.

\textit{Quantum RD Keldysh action.} The dynamics of the considered Fermi gas in $d$ spatial dimensions is governed by the quantum master equation \cite{lindblad1976generators,gorini1976completely,breuer2002,gardiner2004quantum} with Lindblad map $\mathcal{L}$
\begin{equation}
\label{eq:master}
    \dot{\rho}(t)=\mathcal{L}[\rho(t)]=-\frac{i}{\hbar}[H,\rho(t)]+\mathcal{D}[\rho(t)].
\end{equation}
The Hamiltonian $H$ describes coherent free motion in the presence of an external trapping potential $V(\vec{x})$:
\begin{equation}
    H=\int d^d\vec{x} \,  \psi^\dagger(\vec{x})[-J\nabla^2+V(\vec{x})]\psi(\vec{x}).
\label{eq:Hamiltonian_diff_potential}
\end{equation}
Here $J=\hbar^2/(2m)$, while $\psi$, $\psi^\dagger$ are fermionic field operators satisfying canonical anticommutation relations $\{\psi(\vec{x}),\psi^\dagger(\vec{x}')\}\,=\,\delta(\vec{x}-\vec{x}')$. It is important to note that Eq.~\eqref{eq:Hamiltonian_diff_potential} describes coherent-ballistic motion, differently from the classical case which features diffusion, see Fig.~\ref{fig:diffusion}(a). The dissipator $\mathcal{D}[\rho]$ embodies irreversible reaction processes
\begin{align}
    \mathcal{D}[\rho(t)]=\sum_{\alpha}\int d^d\vec{x}\,\Big[\,& L_\alpha(\vec{x})\rho(t)L^\dagger_\alpha(\vec{x})\nonumber\\
    &-\frac{1}{2}\{\,L^\dagger_\alpha(\vec{x})L_\alpha(\vec{x}),\rho(t)\}\Big],
\end{align}
with $\alpha=1,2\dots d$. We focus on binary annihilation reactions $A+A\,\rightarrow\,\emptyset$, see Fig.~\ref{fig:diffusion}(a), modeled by the jump operators
\begin{equation}
L_\alpha(\vec{x})=\sqrt{\Gamma}\,\psi(\vec{x})\, \partial_{x_\alpha}\psi(\vec{x}).
\label{eq:annihilation_continuum_limit}
\end{equation}
The constant $\Gamma$ (units: $\mbox{length}^{d+2}/\mbox{time}$) characterizes the annihilation reactions. In the Supplemental Material \cite{SM}, the jump operator \eqref{eq:annihilation_continuum_limit} is obtained by taking the continuum limit of nearest-neighbors annihilation. The latter is the natural annihilation decay to consider for fermionic particles, where on-site reactions are forbidden. 

Nonequilibrium universal behavior manifests in the power-law decay of the density $n(x,t)=\braket{\psi^{\dagger}(\vec{x}) \psi(\vec{x})}_t$ in time. Power-law decay is a general consequence of the nonlinearity of the binary annihilation process \eqref{eq:annihilation_continuum_limit}, and it can therefore be present both for ballistic (as in the case of \eqref{eq:Hamiltonian_diff_potential}) and diffusive transport of particles. Here, we characterize this decay in the reaction-limited regime of weak dissipation. This regime amounts to considering weak dissipative perturbations \eqref{eq:annihilation_continuum_limit} $\sim\Gamma$ to the integrable (noninteracting) Hamiltonian \eqref{eq:Hamiltonian_diff_potential}. In particular, we take $\hbar n(x,t) \Gamma/J \ll 1$, so that reactions are weak and the density slowly changes in time. We further allow for weak spatial inhomogeneities due the presence of the trapping potential $V(\vec{x}\, \Gamma)$, which we assume to vary on macroscopic length scales $\vec{x} \sim \Gamma^{-1}$. In this limit, the quasiparticle dispersion relation $\epsilon_{k}(\vec{x})=J\vec{k}^2+V(\vec{x}\,\Gamma)$ of $H$ in Eq.~\eqref{eq:Hamiltonian_diff_potential}, with $\vec{k}$ the momentum, is locally modified by the external potential.

In this regard, we study the quantum reaction-limited regime in the Euler-scaling limit \cite{spohn2012large,Dojon20,essler2022short,lossth5,de2022correlation,Durnin2021,Moyal6,GHDfield,DoyonCorr2018,MollerCorr,PerfettoEulerFluct,BMFTlong}. The Euler scale is the largest scale of hydrodynamics, where space-time observation points are large keeping their ratio finite: $\vec{x},t \to \infty$, $\Gamma \to 0$ with $\bar{\vec{x}}=\Gamma \vec{x}$ and $\bar{t}=\Gamma t$ fixed. In the different context of Hamiltonian integrability-breaking perturbations, similar scaling limits have been studied in Refs.~\cite{tGGE2,Durnin2021}. In the ensuing ``Boltzmann regime'', it has been shown \cite{tGGE2} that the slow dynamics of the weakly broken charges of the unperturbed Hamiltonian is governed by the instantaneous GGE of the integrable Hamiltonian. In the context of dissipative systems, in the Euler-scaling limit the similar TGGE assumption has been put forward \cite{tGGE1,tGGE3,tGGE4}, but a derivation of that is missing. In this Letter, we aim at connecting the TGGE assumption to diagrammatics techniques showing which assumptions in the latter eventually allow us to reobtain the former.

To do this, we exploit the Keldysh quantum field theory description \cite{kamenev2011, altland2010,sieberer2016,thompson2023field,tonielli2016,sieberer2023universality} of the open quantum RD dynamics \eqref{eq:master}-\eqref{eq:annihilation_continuum_limit} \cite{SM}. A general feature of Keldysh field theory is the doubling of the $\psi,\,\bar{\psi}$ fields into four fields---$\psi_+,\,\bar{\psi}_+$ and $\psi_-,\,\bar{\psi}_-$---evolving along a ``forward'' ($+$) and a ``backward'' ($-$) contour of the considered time interval, respectively, as sketched in Fig.~\ref{fig:diffusion}(b).
\enquote{Plus} and \enquote{minus} fields are usually rewritten in terms of Keldysh-rotated fields \cite{keldysh1965,SM, larkin1975} $\phi_1,\,\bar{\phi}_1,\,\phi_2,\,\bar{\phi}_2$.
The Keldysh partition function $Z(t)\!=\!\mathrm{tr}[\rho(t)] \!=\! \int \mathbf{D}[\phi_1,\bar{\phi}_1,\phi_2,\bar{\phi}_2]\,\mathrm{exp}\{iS[\phi_1,\bar{\phi}_1,\phi_2,\bar{\phi}_2]\}$
includes full information on the system's microscopic dynamics. The action $S=S_0+S_{\mathcal{D}}$ is composed of two sectors. A quadratic sector $S_0$ describes coherent motion \eqref{eq:Hamiltonian_diff_potential}:
\begin{equation}
    S_0=\int d^dx \, dt'\,[\bar{\phi}_{1}(G_0^R)^{-1}\phi_{1}+\bar{\phi}_{2}(G_0^A)^{-1}\phi_{2}\,],
\label{eq:diffusion_action}
\end{equation}
with $(G_0^{R/A})^{-1}\,=\,i\partial_{t'}+(J\nabla^2-V(\vec{x}))/\hbar\pm i\delta$ the inverse retarded/advanced bare propagator, respectively. The bare Keldysh Green's function $G_0^{K}\equiv -i\braket{\phi_1\bar{\phi}_2}$ associated to $S_0$ is a regularization factor. 
The retarded/advanced propagator $G_0^{R/A}$ (with $V(\vec{x})=0$) in $S_0$ has a structure similar to the classical RD quadratic counterpart  \cite{peliti1985,peliti1986renormalisation, Lee1994, cardy1996, tauber2005,tauber2002dynamic, tauber2014critical}, the difference between the two being in the ballistic quantum motion of $S_0$ compared to the classical diffusive one. The second part $S_{\mathcal{D}}$ of the action contains the interaction vertices of the theory due to annihilation reactions \eqref{eq:annihilation_continuum_limit}:
\begin{align}
    S_{\mathcal{D}} &=\frac{i\Gamma}{4}\int d^dx \, dt'  \Big[ 2(\vec{\nabla}\Bar{\phi}_1\Bar{\phi}_2 + \vec{\nabla}\Bar{\phi}_2\Bar{\phi}_1)\cdot(\phi_1\vec{\nabla}\phi_2+ \nonumber \\
    &+\phi_2\vec{\nabla}\phi_1)+(\vec{\nabla}\Bar{\phi}_1\Bar{\phi}_2 + \vec{\nabla}\Bar{\phi}_2\Bar{\phi}_1)\cdot
    (\phi_1\vec{\nabla}\phi_1+\phi_2\vec{\nabla}\phi_2) -\nonumber \\
    &- (\vec{\nabla}\Bar{\phi}_1\Bar{\phi}_1 + \vec{\nabla}\Bar{\phi}_2\Bar{\phi}_2)\cdot
    (\phi_1\vec{\nabla}\phi_2+\phi_2\vec{\nabla}\phi_1)\Big]\,.
\label{eq:interaction_vertex}    
\end{align}
All the interaction vertices are quartic in the fields, differently from the classical RD field theory, where both cubic and quartic interaction vertices are present. Furthermore, spatial gradients of the fields appear as a consequence of fermionic statistics. 

\textit{Kinetic equation.} In the reaction-limited regime $\hbar n \Gamma/J\ll 1$, the dressed Green's functions $\hat{G}$ can be rewritten in terms of the bare Green's functions $\hat{G}_0$ as the interaction vertices \eqref{eq:interaction_vertex} are expanded perturbatively around the quadratic sector \eqref{eq:diffusion_action}. The sum of all internal one-particle-irreducible contributions to the Feynman diagrams results in the entries of the ``self-energy'' Keldysh-space matrix $\hat{\Sigma}$ \cite{schwinger1961brownian, keldysh1965,kamenev2011, altland2010,sieberer2016,tonielli2016,thompson2023field,sieberer2023universality,buchhold2015kinetic,Duval2023}. The ensuing Dyson equation is pictorially shown in Fig.~\ref{fig:diffusion}(c). 

The Keldysh component of the Dyson equation for $G^K(\vec{x}_1,t_1,\vec{x}_2,t_1)$ determines the kinetic equation \cite{SM}. The large-scale universal limit of this equation is best extracted by performing a Fourier transform in the relative space [time] variable $\vec{x}'=\vec{x}_1-\vec{x}_2$ [$t'=t_1-t_2$], $G^K(\vec{x},t,\vec{k},\epsilon)$  \footnote{In the whole Letter, for lightness of notation, we use the same symbol for a function and its Wigner transform. The two functions can be distinguished from the corresponding arguments.}, i.e., the so-called ``Wigner transform''\cite{weyl1931theory,WignerRev,case2008wigner}. The set of Wigner center-of-mass coordinates $\vec{x}\,=\,(\vec{x_1}+\vec{x_2})/2$ [$t=(t_1+t_2)/2$], as well as momentum $\vec{k}$ and energy $\epsilon$, characterize the effective macroscopic evolution of Green's functions.

\begin{figure}[t]
    \centering
    \includegraphics[width=0.75\columnwidth]{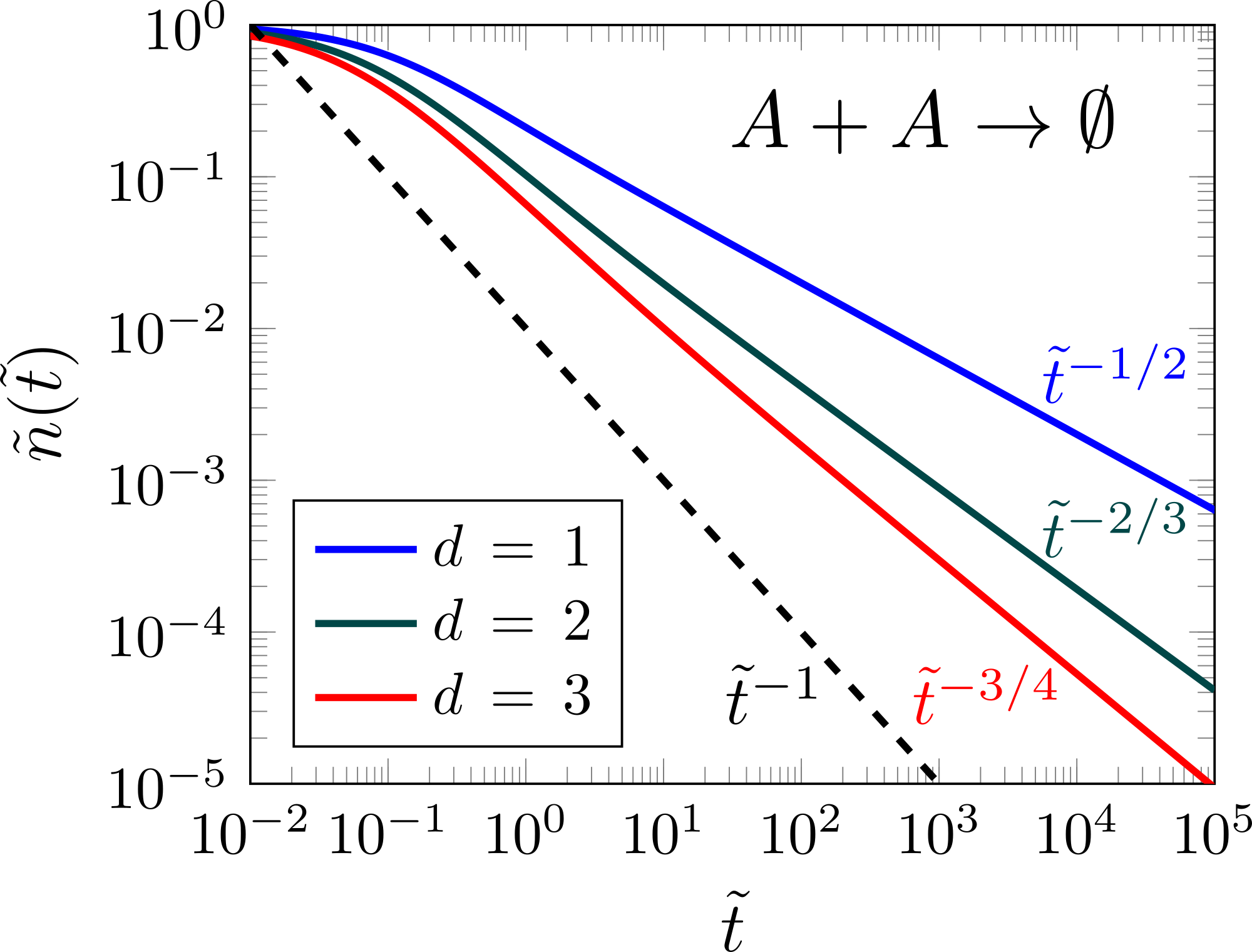}
    \caption{Binary annihilation decay in $d$ dimensions. Solution of the homogeneous Boltzmann equation \eqref{eq:boltzmann} from the Fermi-sea initial state at density $n_0$. The rescaled density $\tilde{n}=n/n_0$ decays algebraically as a function of the dimensionless time $\tilde{t}=n_0^{1+2/d}\Gamma t$. From top to bottom, algebraic decay $\tilde{n}\sim \tilde{t}^{-\frac{d}{d+1}}$ in $d=1,2,3$ (solid lines). The dashed line represents the mean-field decay exponent $\tilde{t}^{-1}$ asymptotically valid in infinite $d$. 
    }
    \label{fig:homo}
\end{figure}

\begin{figure*}[t]
\centering
\includegraphics[width=2\columnwidth]{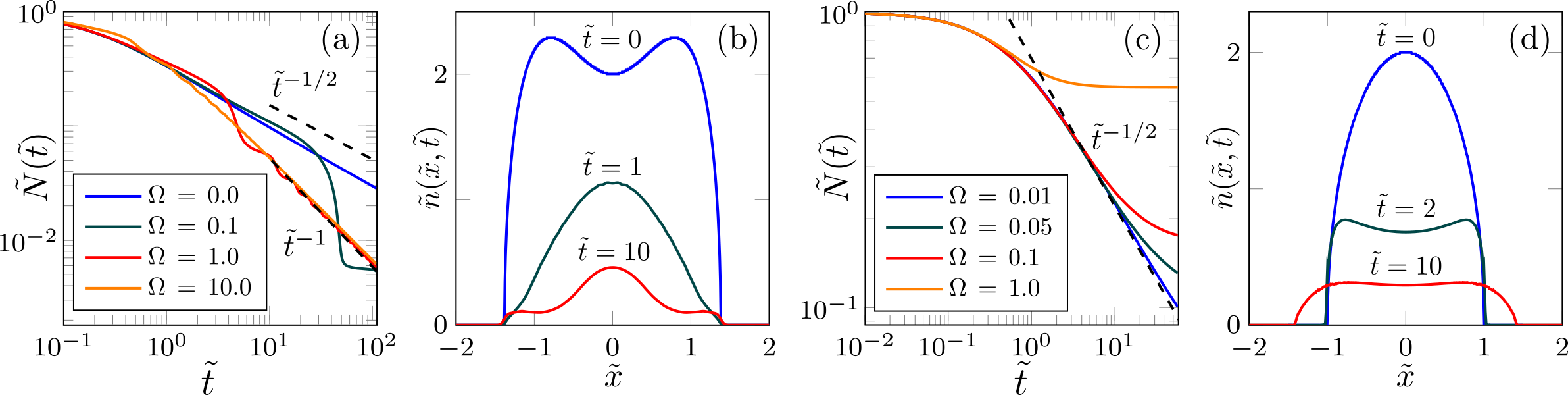}
\caption{Binary annihilation inhomogeneous decay in $d=1$. (a) Double-well to harmonic confinement potential quench: plot of the rescaled particle number $\tilde{N}(\tilde{t})$ as a function of rescaled time $\tilde{t}$ for increasing values of the parameter $\Omega=2n(0,0)[2J/(8\hbar \omega  N_0^3)]^{1/2}$ (from top to bottom). The algebraic time decay gets accelerated as $\Omega$ is increased towards the mean-field prediction $\tilde{N}(\tilde{t})\sim t^{-1}$ (bottom dashed line). 
(b) The corresponding rescaled spatial density $\tilde{n}(\tilde{x},\tilde{t})$ profiles at increasing times (from top to bottom) are plotted as a function of space $\tilde{x}$, with $\Omega=0.1$. For both plots $C=0.8$ and $B=1$. (c) Trap release quench: particle number decay $\tilde{N}(\tilde{t})$ versus time $\tilde{t}$ for increasing $\Omega$ (from bottom to top). A decay exponent $\tilde{N}(\tilde{t})\sim \tilde{t}^{-\xi}$ is approximately observed at intermediate times, with $\xi$ decreasing with $\Omega$, and $\xi=1/2$ for $\Omega=0$ (black dashed line). At longer times a non-algebraic slow decay sets in.
(d) The associated rescaled spatial density $\tilde{n}(\tilde{x},\tilde{t})$ profiles are reported at increasing times (from top to bottom), with $\Omega=0.1$.}
\label{fig:plots}
\end{figure*}

At this point two assumptions are needed: (i) Euler-scaling limit, so slow space-time variations  so as we perform the Moyal-derivative expansion of hydrodynamics  \cite{Moyal1,Moyal2,Moyal3,Moyal5,Moyal6,essler2022short} at leading order in space-time derivatives; (ii) stable quasiparticle excitations, which in the Keldysh formalism translates into a sharply peaked $G^K(\vec{x},t,\vec{k},\epsilon)$ around $\epsilon_{k}(\vec{x})$, as in Fig.~\ref{fig:diffusion}(c). Both (i) and (ii) rely on the weak dissipative integrability breaking $\Gamma \to 0$ and this is why the kinetic equation eventually matches the TGGE prediction.  
This is explicitly shown via the exact relation for the equal-time Keldysh Green's function $G^K(\vec{x},t,\vec{k},t)$ as:
\begin{equation} 
\label{eq:GK}
    iG^K(\vec{x},\vec{k},t,t)\,=\,1\,-\,2\,n(\vec{x},\vec{k},t).
\end{equation}
Here, $n(\vec{x},\vec{k},t)$ is the one-body Wigner function, i.e., the semiclassical phase-space $(\vec{x},\vec{k})$ occupation function \cite{wigner3F,wigner1F,wigner2F}. Within the conditions (i) and (ii), the Wigner function can be identified as the emergent degree of freedom, which obeys the quantum Boltzmann-like equation ($\bar{t}=\Gamma t$ and $\bar{\vec{x}}=\Gamma \vec{x}$)
\begin{equation} 
\label{eq:boltzmann}
\begin{split}
    \Big[\,\partial_{\bar{t}} &+ \vec{v}_g(\vec{k}) \cdot \vec{\nabla}_{\bar{x}}-\vec{\nabla}_{\bar{x}} V/\hbar\cdot \vec{\nabla}_k \,\Big]n(\bar{\vec{x}},\vec{k},\bar{t}) =\\
    &=-\int \frac{d^dq}{(2\pi)^d}(\vec{k}-\vec{q})^2 \,n(\bar{\vec{x}},\vec{k},\bar{t})\,n(\bar{\vec{x}},\vec{q},\bar{t}).
\end{split}
\end{equation}
Crucially, in Euler-scaling limit, we find that $\hat{\Sigma}$ contributes via purely imaginary terms, which determine the r.h.s., named collision integral. The latter is computed in terms of tadpole Feynman diagrams (depicted in Fig.~\ref{fig:diffusion}(c)) at lowest order in the derivatives $\vec{\nabla}$.
The appearing factor $(\vec{k}-\vec{q})^2$ stems from the fermionic statistics. Moreover, the dispersion relation $\epsilon_k(\vec{x})$ ($\vec{v}_g(\vec{k})\,=\,2 J \vec{k}/\hbar$ is the group velocity) is not renormalized. This is a consequence of the integrability-breaking term being purely dissipative. When additional Hamiltonian integrability-breaking perturbations are introduced, the quasiparticle spectrum and the potential $V$ can get possibly renormalized. In $d=1$, the r.h.s. of Eq.~\eqref{eq:boltzmann} has the same form as the one derived in Refs.~\cite{lossth1,Rosso2022,perfetto22,perfetto23,lossth11} assuming the systems relaxes to a TGGE state in between consecutive reactions. This analysis therefore shows how systems whose integrability is weakly broken due to dissipation can be equivalently studied via Keldysh diagrammatic methods. At the same time, it allows us to consider higher dimensional systems.

\textit{Homogeneous decay in $d$ dimensions.} For homogeneous initial states and no trapping potential $V(\vec{x})=0$, the Wigner function $n(\vec{x},\vec{k},t)$ reduces to the momentum-occupation function $n(\vec{k},t)$. We consider a Fermi-sea initial state, with equally populated modes up to some Fermi momentum and a total initial density $n_0$. The asymptotics of the particle density decay can be worked out for generic $d$ \cite{SM}. 
It is convenient to introduce the adimensional rescaled density $\tilde{n}\,=\,n/n_0$ and time $\tilde{t}\,=\,n_0^{1+2/d}\Gamma t$. The long-time asymptotics for $\tilde{n}(\tilde{t})$ is given by the power law
\begin{equation}
\label{eq:homo}
    \tilde{n}(\tilde{t})\sim \Bigg[\,\frac{[(\alpha_d \, \Theta_d) (d-2)!!]^2}{2^d (d+1)^d (2\pi)^{2d}}\,\Bigg]^{\frac{1}{d+1}}\tilde{t}^{-\frac{d}{d+1}},
\end{equation}
with $\Theta_d$ the $d$-dimensional solid angle \footnote{$\alpha_d=1$ for $d$ even, $\alpha_d=\sqrt{\pi/2}$ for $d$ odd.}. The solution of the homogeneous Boltzmann equation for $d=1,2,3$ from the Fermi-sea initial state is shown in Fig.~\ref{fig:homo}. In $d=1$, the decay exponent is $1/2$ in agreement with the TGGE prediction \cite{lossth1,Rosso2022,perfetto22, perfetto23,lossth11}. We find that the algebraic decay is different from the mean-field $\tilde{t}^{-1}$ even for $d>1$ and it approaches the latter only for $d\to \infty$. This result is surprising and fundamentally different from the classical $A+A\to\emptyset$ RD dynamics. Therein, non mean-field algebraic decay is possible only in $d=1$ in the diffusion-limited regime $\hbar \Gamma n/J \sim 1$ \cite{peliti1985,peliti1986renormalisation, Lee1994, cardy1996,tauber2005,tauber2002dynamic, tauber2014critical}, as a consequence of spatial fluctuations. Conversely, the different dimensional dependence of the exponent in Eq.~\eqref{eq:homo} does not emerge due to spatial fluctuations but from the universal large-scale limit \eqref{eq:boltzmann} of the quantum dynamics. We remark that the power-law decay \eqref{eq:homo} is dictated by the nonlinearity of the binary annihilation reaction. In addition the decay \eqref{eq:homo} beyond mean field is not specific of the zero temperature initial Fermi-sea state, since it also describes the dynamics ensuing from finite temperature initial states, with the temperature changing the amplitude but not the exponent of the decay \cite{SM}.

\textit{Inhomogeneous decay in one dimension.} 
We now consider a one-dimensional quantum quench of a slowly varying trapping potential from the prequench $V_0(\varepsilon x)=A(\varepsilon x)^4/4 - m\omega^2 (\varepsilon x)^2/2$ double-well to the postquench harmonic $V(\varepsilon x)=m\omega^2 (\varepsilon x)^2/2$ form. The small adimensional parameter $\varepsilon=\hbar n(0,0) \Gamma/J$ ensures the potentials to be slowly varying in $x$. This analysis is thus an example of the application of generalized hydrodynamics to the case of a weakly varying external field \cite{GHDfield}, with the additional presence here of slow dissipation. A similar setting has been consider in Ref.~\cite{Rosso2022}, where also the prequench potential $V_0(x)$ is harmonic. The initial condition is the local-density approximation of the ground state of the Hamiltonian $H$ in Eq.~\eqref{eq:Hamiltonian_diff_potential} with potential $V_0(x)$. We set the initial particle number to $N_0$ and we perform Euler scaling according to the parameters of the harmonic potential $V(x)$: for the particle number $\tilde{N}=N/N_0$, time $\tilde{t}=\varepsilon t (2N_0)^{3/2}J/(\ell_{HO}^3\hbar n(0,0))$, space $\tilde{x}= \varepsilon x/(\sqrt{2N_0}\ell_{HO})$, momentum $\tilde{k}=k \ell_{HO}/\sqrt{2N_0}$ and density $\tilde{n}=n (2\pi \ell_{HO}/\sqrt{2 N_0})$, with $\ell_{HO}=\sqrt{\hbar/(m\omega)}$. The adimensional parameter $\Omega=2n(0,0)[2J/(8\hbar \omega  N_0^3)]^{1/2}$ quantifies the interplay between coherent motion ($J$) and confinement ($\omega$). In the rescaled phase space $(\tilde{x},\tilde{k})$, the initial state is $n(\tilde{x},\tilde{k},t)=1$ if $B-\tilde{k}^2+\tilde{x}^2 -C \tilde{x}^4>0$, and zero otherwise. Here $C=A \hbar N_0/(m^2 \omega^3)$ and $B=\mu/(\hbar N_0 \omega)$, with the chemical potential $\mu$ fixing the initial particle number $N_0$.

In Fig.~\ref{fig:plots}(a), the decay of $\tilde{N}$ as a function of $\tilde{t}$ is shown. The decay accelerates periodically as $\Omega$ is increased since particles bounce off the potential walls and gather up at the center of the well, as shown in Fig.~\ref{fig:plots}(b) for the density $\tilde{n}(\tilde{x},\tilde{t})$. As a consequence of such breathing motion all decay profiles converge from the short-time $\tilde{N}(\tilde{t})\sim \tilde{t}^{-1/2}$ algebraic behavior towards the mean-field asymptotic decay $\tilde{N}(\tilde{t})\sim \tilde{t}^{-1}$. 

Next, we consider the long-time decay for the deconfinement dynamics of an harmonic trap-release quench of the Fermi gas from $V_0(x)=m \omega^2 (\varepsilon x)^2/2$ to $V(x)=0$. We rescale also in this case variables with respect to the harmonic potential $V_0(x)$ parameters. In Fig.~\ref{fig:plots}(c) the decay of $\tilde{N}$ in time $\tilde{t}$ is reported. One first observes \cite{SM} an approximate algebraic decay $\tilde{N}(\tilde{t})\sim \tilde{t}^{-\xi}$, with an exponent $\xi$ continuously decreasing as $\Omega$ is increased (from the value $\xi=1/2$ at $\Omega=0$). At longer times, an unexpectedly slow decay, when compared to any power law, sets in. This slow decay is unexpected because it is not solely determined by the decrease of the density due to the expansion in free space, in Fig.~\ref{fig:plots}(d), but also by the fermionic statistics. The most relevant reactions at low densities in the trap-release protocol, indeed, take place between particles with same momenta. The fermionic statistics, manifest in the factor $(\tilde{k}-\tilde{q})^2$ in Eq.~\eqref{eq:boltzmann}, thereby suppresses these reactions and determine the decay of Figs.~\ref{fig:plots}(c)-(d).

\textit{Summary.} 
We provided a Keldysh field-theory description of quantum RD dynamics of binary annihilation $A+A\to\emptyset$. We analytically derived in the Euler-scaling limit the universal large-scale Boltzmann equation in arbitrary dimension $d$ describing the reaction-limited regime of slow reactions $\hbar n \Gamma/J\ll 1$. In $d=1$, our results match the prediction from the TGGE ansatz, connecting the latter to field-theoretical diagrammatic expansions. For homogeneous systems, we analytically showed that the density algebraic decay exponent features an unexpected dependency on $d$ and it deviates from mean-field value even in $d>1$, in contrast with classical RD dynamics. In one-dimensional inhomogeneous setups involving a trapping potential, we found that the decay is either accelerated towards the mean-field value (confined systems), or severely slowed down (deconfined systems). Our results find a natural application in cold-atomic experiments involving two-body losses \cite{lossexp0,lossth7,lossexpF2,lossexpF3} in the strong-dissipation Zeno regime. From the formulation here proposed, several relevant questions can be addressed. As an example, one can assess the impact of elastic-Hamiltonian collisions on the decay exponent \cite{lossexpF2,lossexpF3,FHth1,FHth2,FHth3}. The presence of Hamiltonian-integrability breaking perturbations can, indeed, result into hydrodynamic diffusion \cite{GHDintbreakDiff}, and it would be interesting to study the possible impact of diffusion on the asymptotic power-law decay of the density. Away from the reaction-limited, it is also crucial to study the quantum diffusion-limited regime $\hbar n \Gamma/J \sim 1$ via renormalization group schemes, as done for the classical RD \cite{cardy1996,tauber2005,tauber2002dynamic,tauber2014critical}.   

\textit{Acknowledgements.} 
We acknowledge fruitful discussion with M. Buchhold, S. Diehl and J.P. Garrahan. F.G. thanks Universit\"at T\"ubingen for hospitality, and acknowledges support from Università di Trento and Collegio Bernardo Clesio. 
G.P. acknowledges support from the Alexander von Humboldt Foundation through a Humboldt research fellowship for postdoctoral researchers. We acknowledge financial support in part from EPSRC Grant no.\ EP/R04421X/1 and EPSRC Grant no.\ EP/V031201/1. We are also grateful for funding from the Deutsche Forschungsgemeinsschaft (DFG, German Research Foundation) under Project No. 435696605, the Research Unit FOR 5413/1, Grant No. 465199066 and the Research Unit FOR 5522/1, Grant No. 499180199.

\bibliography{refs}


\setcounter{equation}{0}
\setcounter{figure}{0}
\setcounter{table}{0}
\renewcommand{\theequation}{S\arabic{equation}}
\renewcommand{\thefigure}{S\arabic{figure}}
\makeatletter
\onecolumngrid
\newpage
\setcounter{secnumdepth}{3}
\pagestyle{plain}

\begin{center}
{\Large SUPPLEMENTAL MATERIAL}
\end{center}
\begin{center}
\vspace{0.8cm}
{\Large Large-scale universality in Quantum Reaction-Diffusion from Keldysh field theory}
\end{center}
\begin{center}
Federico Gerbino,$^{1}$ Igor Lesanovsky,$^{2,3}$ and Gabriele Perfetto$^{2}$
\end{center}
\begin{center}
$^1${\em Laboratoire de Physique Th\'eorique et Modèles Statistiques, Université Paris-Saclay, CNRS, 91405 Orsay, France}\\
$^2${\em Institut f\"ur Theoretische Physik, Universit\"at T\"ubingen, Auf der Morgenstelle 14, 72076 T\"ubingen, Germany}\\
$^3${\em School of Physics and Astronomy and Centre for the Mathematics and Theoretical Physics of Quantum Non-Equilibrium Systems, The University of Nottingham, Nottingham, NG7 2RD, United Kingdom}\\
\end{center}

\setcounter{page}{1}

This Supplemental Material provides additional information on the model and the calculations at the basis of the results presented in the main text. In Sec.~\ref{sec:continuum}, we derive the continuum-space Lindblad dynamics, Eqs.~(1)-(4) of the main text, by taking the continuum limit of the corresponding lattice model. 
In Sec.~\ref{sec:open_keldysh}, we briefly recall the basic aspects of the Keldysh field theory for open quantum systems, which are needed for the understanding of the action formulation in Eqs.~(5) and (6) of the main text. In Sec.~\ref{sec:self_energy_calculations}, we briefly report the main steps underlying the diagrammatic calculation of the self-energy Keldysh matrix in the Euler-scaling limit and the associated derivation of the Boltzmann equation in Eq.~(8) of the main text. In Sec.~\ref{sec:calculations_hom_decay}, we report the derivation of Eq.~(9) of the main text concerning the long-time asymptotic of the density decay for homogeneous systems. In Sec.~\ref{sec:effective_exp_supp}, we compute the effective exponent for the deconfining dynamics ensuing from the trap-release quench of the trapping potential (see Fig.~3 of the main text).
	
\section{Lindblad dynamics in the space-continuum limit} 
\label{sec:continuum}

\begin{figure}
    \centering
    \includegraphics[width=\textwidth]{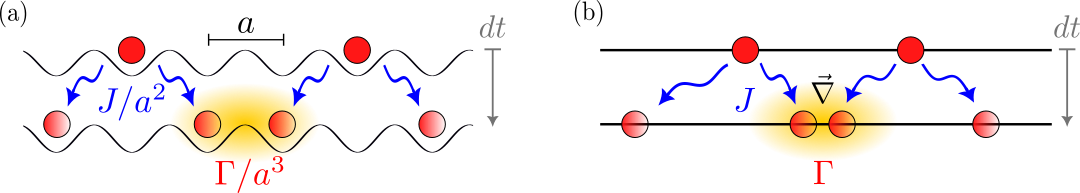}
    \caption{\textbf{Space-continuum limit.}
    (a) Quantum RD dynamics on a one-dimensional lattice with lattice spacing $a$.  Fermions coherently hop (blue-wiggly lines) between neighbouring sites of an optical lattice (black wavy line) at rate $J/(a^2\hbar)$. Binary annihilation $A+A\to\emptyset$ reactions take place between neighbouring sites at rate $\Gamma/a^3$, as sketched in yellow. 
    (b) Taking the continuum limit $a\,\rightarrow\,0$ of the discrete-lattice model, defines the continuous quantum RD dynamics under examination in Eqs.~\eqref{eq:Hamiltonian_continuum_sup} and \eqref{eq:lindblad_d_dimension_continuum_supp}. Here, $J/\hbar$ has units [$\mbox{length}^{2}/\mbox{time}$] and it controls coherent motion, while binary annihilation reactions occur via derivative couplings of the fields with coupling constant $\Gamma$, with units [$\mbox{length}^{3}/\mbox{time}$].}
    \label{fig:continuum}
\end{figure}

We start by introducing, for the sake of illustrative purposes, a one-dimensional lattice model, which is pictorially represented in Fig.~\ref{fig:continuum}(a). Let us consider an infinite chain whose sites labelled by index $j\in\mathbb{Z}$ can be either occupied by one fermion $n_j\ket{...\bullet_j...}=\ket{...\bullet_j...}$ or empty $n_j\ket{...\circ_j...}=0$. The number operator $n_j=c_j^{\dagger}c_j$ is written in terms of fermionic destruction ($c_j$) and creation ($c_j^{\dagger}$) operators satisfying the anticommutation relation $\{c_i,c_j^\dagger\}=\delta_{i,j}$. We also introduce the lattice spacing $a$, i.e., the shortest length scale of the problem. The hopping Hamiltonian reads
\begin{equation}
    H\,=\,-\,\frac{J}{a^2}\,\sum_j\,(\,c_{ja}^\dagger c_{(j+1)a}+c^\dagger_{(j+1)a} c_{ja}\,)\,+\,\sum_j\,c^\dagger_{ja} V_{ja} c_{ja}\,.
\label{eq:hopping_Hamiltonian}
\end{equation}
We remark that we here follow the convention of Ref.~\onlinecite{tauber2005} for the hopping amplitude notation so that $J/\hbar$ has units [$\mbox{length}^{2}/\mbox{time}$] of a diffusion constant, while $J/(a^2 \hbar)$ has units [$\mbox{time}^{-1}$] of a rate. We further include the effect of a position-dependent trapping potential $V_{ja}$. In the absence of the latter, the fermionic hopping Hamiltonian \eqref{eq:hopping_Hamiltonian} has been studied in Refs.~\onlinecite{van_horssen2015,perfetto22} for quantum RD models on a chain. The reaction part is encoded into the jump operators of the Lindbladian and it is given by two-body annihilation process $A+A\to\emptyset$. This destroys two neighbouring fermions at a rate $\Gamma/a^3$ ($\Gamma$ has units [$\mbox{length}^{3}/\mbox{time}$]) \begin{equation}
    \dot{\rho}(t)=-i[H,\rho(t)]+\sum_{j} \left[L_{ja}\rho {L_{ja}}^\dagger-\frac{1}{2}\left\{{L_{ja}}^\dagger L_{ja},\rho \right\}\right], \quad \mbox{with} \quad L_{ja}=\sqrt{\frac{\Gamma}{a^{3}}}\,c_{ja} c_{(j+1)a}. 
\label{eq:Lindblad_1d}
\end{equation}
Clearly, for fermions, one cannot define reactions occurring on the same site as this would violate the exclusion principle. The annihilation process between neighbouring sites is sketched in Fig.~\ref{fig:continuum}(a). 

The generalization of Eqs.~\eqref{eq:hopping_Hamiltonian} and \eqref{eq:Lindblad_1d} to generic space dimension $d$ can be simply accomplished by introducing a $d-$dimensional hypercubic lattice. Lattice points on this lattice are identified by a $d-$dimensional vector $\textbf{j}=(j_1,\dots j_{\alpha},\dots j_d)$ of integer numbers $j_{\alpha}\in\mathbb{Z}$ ($\alpha=1,2\dots d$). Then one replaces $c_{ja}$ and $c_{(j+1)a}$ (the same applying for the corresponding creation operators) with $c_{\textbf{j}a}$ and $c_{\textbf{j}a+a\textbf{e}_\alpha}$, $\textbf{e}_\alpha$ being the unit-vector pointing to direction $\alpha\,=\,1,...,d$. The Hamiltonian and the Lindblad master equation then read 
\begin{equation}
    H=-\,\frac{J}{a^2}\,\sum_{\alpha=1}^d\,\sum_{\textbf{j}}\,(\,c_{\textbf{j}a}^{\dagger}c_{\textbf{j}a+a\textbf{e}_\alpha}+c_{\textbf{j}a+a\textbf{e}_\alpha}^{\dagger}c_{\textbf{j}a})\,+\,\sum_{\alpha=1}^d\,\sum_{\textbf{j}}\,c_{\textbf{j}a}^\dagger V_{\textbf{j}a} c_{\textbf{j}a},
\label{eq:Hamiltonian_hopping_d_dimension}
\end{equation}
and
\begin{equation}
\dot{\rho}(t)=-i[H,\rho(t)]+\sum_{\alpha=1}^{d}\sum_{\textbf{j}} \left[L_{\textbf{j}a}^{(\alpha)}\rho {L_{\textbf{j}a}^{(\alpha)}}^\dagger-\frac{1}{2}\left\{{L_{\textbf{j}a}^{(\alpha)}}^\dagger L_{\textbf{j}a}^{(\alpha)},\rho \right\}\right], \quad \mbox{with} \quad
L_{\textbf{j}\alpha}^{(\alpha)}=\sqrt{\frac{\Gamma}{a^{d+2}}}\,c_{\textbf{j}a} c_{\textbf{j}a+a\textbf{e}_\alpha}\,.
\label{eq:lindblad_d_dimension}
\end{equation}
Equation \eqref{eq:Hamiltonian_hopping_d_dimension} shows that particles can hop isotropically (with the same amplitude) to one of the neighbouring sites in any direction $\alpha$. Annihilation reactions, according to Eq.~\eqref{eq:lindblad_d_dimension}, also take place isotropically.

Let us take the continuum limit of this model defined by the substitutions $\textbf{j}a\,\rightarrow\,\vec{x}$,  $\sum_\textbf{j}\,\rightarrow\,\int\,d^dx/a^d$ and by considering $a\,\rightarrow\,0$.
We may then write:
\begin{equation}
    \lim_{a\rightarrow 0}\,\frac{L_{\textbf{j}a}^{(\alpha)}}{a^{d/2}}=\lim_{a\rightarrow 0}\,\sqrt{\frac{\Gamma}{a^d}}\,\frac{c_{\textbf{j}a}c_{\textbf{j}a+a\textbf{e}_\alpha}}{a^{d/2+1}}=\lim_{a\rightarrow 0}\,\sqrt{\frac{\Gamma}{a^d}}\,\frac{c_{\textbf{j}a}}{a^{d/2}}\,\frac{c_{\textbf{j}a+a\textbf{e}_\alpha}-c_{\textbf{j}a}}{a}.
\label{eq:continuum_limit_L_d_dimensions}
\end{equation}
Introducing the field operators
\begin{equation}
\psi(\vec{x})=\frac{c_{\textbf{j}a}}{a^{d/2}}, \quad \mbox{with} \quad \{\psi(\vec{x}),\psi^{\dagger}(\vec{x}')\}=\delta(\vec{x}-\vec{x}'),
\label{eq:field_operators}
\end{equation}
one recognises that Eq.~\eqref{eq:continuum_limit_L_d_dimensions} is the definition of a partial derivative in direction $x_\alpha$, namely:
\begin{equation}
    \lim_{a\rightarrow 0}\,\frac{L_{\textbf{j}a}^{(\alpha)}}{a^{d/2}}=\sqrt{\Gamma}\,\psi(\vec{x})\,\frac{\partial \psi}{\partial x_\alpha}=L_{\alpha}(\vec{x})\,.
\label{supeq:continuum_limit_ann_jump}
\end{equation}
Repeating the same procedure for the Hamiltonian \eqref{eq:Hamiltonian_hopping_d_dimension}, the definition of a second partial derivative $\frac{\partial^2}{\partial x_\alpha^2}$ appears
\begin{equation}
    H=\int\,d^dx\,\psi^\dagger(\vec{x})\,[\,-J\nabla^2+V(\vec{x})\,]\,\psi(\vec{x}).
\label{eq:Hamiltonian_continuum_sup}
\end{equation}
This equation is readily recognised as Eq.~(2) of the main text. For the dissipator in Eq.~\eqref{eq:lindblad_d_dimension}, we also have 
\begin{equation}
\dot{\rho}(t)=-i[H,\rho(t)]+\Gamma\,\int\,d^dx\,\Big[\,\psi\vec{\nabla}\psi\,\rho\,\cdot\,\vec{\nabla}\psi^\dagger\psi^\dagger\,-\,\{\,\vec{\nabla}\psi^\dagger\psi^\dagger\,\cdot\,\psi\vec{\nabla}\psi,\,\rho\,\}\,\Big].
\label{eq:lindblad_d_dimension_continuum_supp}
\end{equation}
Writing the dot product $\cdot$ explicitly in terms of the associated $\alpha=1,2\dots d$ components, Eq.~\eqref{eq:lindblad_d_dimension_continuum_supp} is recognised as Eq.~(3) of the main text. In Fig.~\ref{fig:continuum}(b), we pictorially show the continuum formulation of the model (in one dimension for the sake of simplicity in the illustration).

\section{Keldysh path integral formulation of open systems}
\label{sec:open_keldysh}
We report here the main steps underlying the construction of Keldysh field theory for dissipative many-body quantum systems \cite{altland2010,sieberer2016,tonielli2016, kamenev2011,thompson2023field}. We then report the main definitions concerning the associated Green functions, which are necessary to understand the derivation of the quantum Boltzmann equation in the main text.

Let us consider the formal solution to a generic quantum Master equation of the form, e.g., in Eq.~\eqref{eq:lindblad_d_dimension}. We start by considering a zero-dimensional system with no spatial structure (no sum over lattice sites $\mathbf{j}$). We can Trotter-decompose the time-evolution operator in infinitely many time slices, thus writing:
\begin{equation}
    \rho(t)= e^{\mathcal{L}t}[\rho(0)]=\lim_{N\rightarrow \infty}\,(\,1+\mathcal{L}\delta t\,)^N[\rho(0)],
\end{equation}
with $\delta t\,=\,t/N$. Hence, at any time instant $t_n\,=\,n\delta t$, with $n=1,2\dots N$, the subsequent density matrix operator evolves via:
\begin{equation}
    \rho_{n+1}=(\,1+\mathcal{L}\delta t\,)[\rho_n]+\,O(\delta t^2).
\label{sup:step_1_path}
\end{equation}
It is now possible to introduce two sets of fermionic coherent states, labelled by Grassmann fields $\psi_n^{\pm}$ which are eigenvalues of the fermionic destruction operator. From the resolution for the identity operator one has:
\begin{equation}
    \rho_n=\int\,d\psi_n^+\,d\bar{\psi}_n^+ d\psi_n^-\,d\bar{\psi}_n^-\,e^{-\bar{\psi}^+_n \psi^+_n -\bar{\psi}^-_n \psi^-_n}\,\bra{\psi_n^+}\rho_n\ket{-\psi_n^-}\,\ket{\psi_n^+}\bra{-\psi_n^-}.
\label{sup:step2_path}
\end{equation}
We label $\ket{\psi_n^+}$ the states acting at time slice $n$ on the density operator from the left, hence evolving the latter in the forward direction along the so-called forward (+) contour. Conversely, states $\ket{\psi_n^-}$ act on the density operator from the right and thus evolve backward in time along a backward (-) contour (cf. the central panel of Fig.~(1) of the main text). 
Doubling the field variables from two to four fields evolving along anti-parallel time axes is a peculiarity of out-of-equilibrium quantum field theories \cite{altland2010,sieberer2016,tonielli2016, kamenev2011,thompson2023field}. Besides, an additional minus sign in front of $\ket{-\psi_n^-}$ and $\bra{-\psi_n^-}$ coherent states has been added with no effect on the result of the calculation.
Inserting Eq.~\eqref{sup:step2_path} into \eqref{sup:step_1_path}, we arrive to:
\begin{align}
    \bra{\psi_{n+1}^+}\rho_{n+1}\ket{-\psi_{n+1}^-}=\int\,d\psi_n^+d\bar{\psi}_n^+\,&d\psi_n^-d\bar{\psi}_n^-\,e^{-\bar{\psi}^+_n \psi^+_n -\bar{\psi}^-_n \psi^-_n} \nonumber\\
    &\bra{\psi_n^+}\rho_n\ket{-\psi_n^-}\,
    \bra{\psi_{n+1}^+}\big\{ (1+\mathcal{L}\delta t)\left[\ket{\psi_{n}^+}
    \bra{-\psi_n^-}\right]\big\}\ket{-\psi_{n+1}^-}\,+\,O(\delta t^2).
\end{align}
Evaluation of the matrix element in squared brackets, together with the normalisation factors of coherent states completeness relations $e^{-\bar{\psi}^+_n \psi^+_n -\bar{\psi}^-_n \psi^-_n}$,  leads to:
\begin{align}
    \bra{\psi_{n+1}^+}\rho_{n+1}\ket{-\psi_{n+1}^-}=\int\,d\psi_n^+d\bar{\psi}_n^+\,&d\psi^-_n d\bar{\psi}^-_n
    \,\bra{\psi_n^+}\rho_n\ket{-\psi_n^-} \nonumber \\
    &\mathrm{exp}\Big\{ i\delta t\Big[ -i\partial_t\bar{\psi}^+_n\psi^+_n
    -i\bar{\psi}^-_n\partial_t\psi^-_n -i\mathcal{L}[ \psi_n^+,\bar{\psi}_{n+1}^+,\psi_{n+1}^-,\bar{\psi}_n^-]\Big]\Big\}\,+\,O(\delta t^2).
\label{supeq:path_integral_3}
\end{align}
In the previous equation, the function $\mathcal{L}[ \psi_n^+,\bar{\psi}_{n+1}^+,\psi_{n+1}^-,\bar{\psi}_n^-]$ is the Lindbladian evaluated in terms of Grassmann fields after the Hamiltonian and the jump operators have been normally ordered so that creation operators always lie to the left of destruction ones. We also introduced the notation $\partial_t \psi_n=(\psi_{n+1}-\psi_n)/\delta t$. 

Repeated iteration of the previous equation from time slice $n\,=\,1$ to $n\,=\,N$ connects the initial time to the final time density matrix. One can then take the limit $N\,\rightarrow\,\infty$, thus neglecting the $O(\delta t^2)$ terms and performing the double substitution $\sum_n \delta t\,\rightarrow\,\int dt$, $\prod_n d\psi_n\,\rightarrow\,\mathbf{D}\psi$. The trace operation joins the forward and backward contour leading to the Keldysh partition function:
\begin{equation}
    Z(t)= \int \mathbf{D}[\psi_+,\bar{\psi}_+,\psi_-,\bar{\psi}_-]\,\bra{\psi_+(0)}\rho(0)\ket{-\psi_-(0)}\,e^{iS[\psi_+,\bar{\psi}_+,\psi_-,\bar{\psi}_-]}\,.
\end{equation}
In the following, we will neglect the boundary term $\bra{\psi_+(0)}\rho(0)\ket{-\psi_-(0)}$ as it does not affect the large-scale Boltzmann equation, which is solely determined by the bulk part of the action. We therefore focus on the bulk part of the action only and we require the associated normalisation constraint
\begin{equation}
    Z(t)=\mathrm{tr}[\rho(t)]= \int \mathbf{D}[\psi_+,\bar{\psi}_+,\psi_-,\bar{\psi}_-]\,e^{iS[\psi_+,\bar{\psi}_+,\psi_-,\bar{\psi}_-]}=1.
\end{equation}
The bulk part of the Keldysh action is denoted as $S$ and it reads
\begin{equation}
    S[\psi_+,\bar{\psi}_+,\psi_-,\bar{\psi}_-]= \int_{0}^{t}\,dt'\, (\, \bar{\psi}_+i\partial_{t'}\psi_+ - \bar{\psi}_-i\partial_{t'}\psi_- - i \mathcal{L}(\psi_+,\bar{\psi}_+,\psi_-,\bar{\psi}_-),
\label{supeq:keldysh_action_1}
\end{equation}
with the Lindbladian:
\begin{equation}
    \mathcal{L}(\psi_+,\bar{\psi}_+,\psi_-,\bar{\psi}_-)\,=\, -i(H_+-H_-)\,+\,\sum_{\alpha}\,\big[\, \bar{L}_{\alpha,-}L_{\alpha,+}-\frac{1}{2}(\bar{L}_{\alpha,+}L_{\alpha,+}+\bar{L}_{\alpha,-}L_{\alpha,-})\,\big].
\label{supeq:keldysh_action_2}
\end{equation}
Functions $H_{\pm}=H(\bar{\psi}_{\pm},\psi_{\pm})$, $L_{\alpha,\pm}=L_{\alpha}(\bar{\psi}_{\pm},\psi_{\pm})$ and $\bar{L}_{\alpha,\pm}=L^{\dagger}_{\alpha}(\bar{\psi}_{\pm},\psi_{\pm})$ 
are evaluated on the forward (+) or backward (-) contour after normal ordering, as explained also after Eq.~\eqref{supeq:path_integral_3}.

Clearly, the field theory described by the preceding Keldysh action applies to zero-dimensional systems without a space structure. However, one can generalise the derivation to spatially-extended systems, with the representation of the identity via a set of coherent states $\ket{\{\psi_n\}}=\ket{\psi_{n,\mathbf{1}}}\ket{\psi_{n,\mathbf{2}}} \dots \ket{\psi_{n,\mathbf{j}}} \dots$ labelled on a time slice $n$ by the lattice site coordinate $\mathbf{1},\mathbf{2}\dots \mathbf{j}\dots$ (the bold face notation representing a vector of integer coordinates with the notation introduced after Eq.~\eqref{eq:Lindblad_1d}):
\begin{equation}
    \mathbb{I}=\int \prod_\textbf{j} d\psi_{n,\textbf{j}} d\bar{\psi}_{n,\textbf{j}} e^{-\sum_\textbf{j} \bar{\psi}_{n,\textbf{j}} \psi_{n,\textbf{j}}} \ket{\{\psi_n\}}\bra{\{\psi_n\}}.
\end{equation}
In this case, the continuum-space limit turns the sum over lattice sites $\textbf{j}$ into a space integral, while it includes in the functional measure $\mathbf{D}[\psi,\bar{\psi}]$ an infinite product over the spatial field configurations evaluated at each time slice. Accordingly, the Keldysh action for the fermionic binary annihilation dynamics, described by the Lindblad equation in Eqs.~\eqref{eq:Hamiltonian_continuum_sup} and \eqref{eq:lindblad_d_dimension_continuum_supp}, in the $\pm$ basis reads:
\begin{align}
        S =S_0+S_{\mathcal{D}}&=\int d^dx dt\, \frac{1}{\hbar}\,\Big[ \,\bar{\psi}_{+}(i\hbar\partial_t +J\nabla^2-V(\vec{x}))\psi_{+} 
        - \bar{\psi}_{-} (i\hbar\partial_t +J\nabla^2-V(\vec{x}))\psi_{-}\,\Big]+ \nonumber \\
        &+ i\Gamma \int  d^dx\,dt \,\Big[\,\frac{1}{2}\vec{\nabla}\bar{\psi}_+\Bar{\psi}_+ \cdot \psi_+\vec{\nabla}\psi_+ + \frac{1}{2}\vec{\nabla}\bar{\psi}_-\Bar{\psi}_- \cdot \psi_- \vec{\nabla}\psi_- -
       \vec{\nabla}\bar{\psi}_-\Bar{\psi}_- \cdot \psi_+\vec{\nabla}\psi_+ \,\Big]\,,
\label{supeq:keldysh_pm}
\end{align}
where the r.h.s. of the second equality on the first line is the quadratic action $S_0$ describing coherent motion in a continuous $d$-dimensional space \eqref{eq:Hamiltonian_continuum_sup}. The quartic action $S_{\mathcal{D}}$ on the second line represents, instead, the interaction vertices due to nearest-neighbour reactions \eqref{supeq:continuum_limit_ann_jump}. We will use henceforth the standard naming ``interaction vertices'' used in field theory for the nonquadratic (in this case quartic) part $S_{\mathcal{D}}$ of the action. By means of the Keldysh rotation of Eq.~\eqref{eq:rotation_K_fermions} which we shall later introduce, the former line gives rise to the quadratic action in Eq.~(5) describing free coherent motion, while the latter generates the dissipative sector discussed in Eq.~(6). The two-point Green's functions associated to the Keldysh action are of central relevance. We list them here as they will be needed in the derivation of the quantum Boltzmann equation in Sec.~\ref{sec:self_energy_calculations}. We use henceforth in the Supplemental Material, the compact 4-vector notation $x=(\vec{x},t)$:
\begin{subequations}
\begin{align}
        & iG^{T}(x_1,x_2)=\langle \psi_+(x_1)\bar{\psi}_+(x_2) \rangle= \braket{T\left[c(x_1)c^{\dagger}(x_2)\right]}, \label{eq:pmgreens_ferm_1} \\
        & iG^<(x_1,x_2)=\langle \psi_+(x_1)\bar{\psi}_-(x_2) \rangle=-\langle c^\dagger (x_2) c(x_1)\rangle,\label{eq:pmgreens_ferm_2}  \\
        & iG^>(x_1,x_2)=\langle \psi_-(x_1)\bar{\psi}_+(x_2) \rangle = \braket{c(x_1)c^{\dagger}(x_2)},\label{eq:pmgreens_ferm_3}\\ 
        & iG^{\tilde{T}}(x_1,x_2)=\langle \psi_-(x_1)\bar{\psi}_-(x_2) \rangle=\braket{\widetilde{T}\left[c(x_1)c^{\dagger}(x_2)\right]}. \label{eq:pmgreens_ferm_4} 
\end{align}
\label{eq:pmgreens}%
\end{subequations}
In Eqs.~\eqref{eq:pmgreens_ferm_1} and \eqref{eq:pmgreens_ferm_4}) the symbols $T$ and $\widetilde{T}$ denote time ordering and anti-time ordering along the Keldysh contour, respectively (the operator evaluated at the larger/smaller time goes to the left, respectively, possibly taking a minus sign if a permutation is needed). Equation \eqref{eq:pmgreens} also shows the direct connection between the Green functions and dynamical correlation functions of creation and destruction operators. We refer the reader to Section $5.2$ of Ref.~\onlinecite{gardiner2004quantum} for a detailed discussion of correlation functions in dissipative systems. Because of probability conservation, the four functions are not independent, i.e., they satisfy the identity:
\begin{equation}
\label{eq:probcons+-}
    G^{T}(x_1,x_2)+G^{\Tilde{T}}(x_1,x_2)- G^< (x_1,x_2)- G^>(x_1,x_2)\,=\,0\,.
\end{equation}
It is then customary to introduce the Keldysh rotation of the fields, which automatically deletes the spurious degrees of freedom, namely, one correlation function is identically vanishing after the rotation. Following the convention by Larkin and Ovchinnikov \cite{larkin1975}, we define:
\begin{equation}
    \phi_1=\frac{\psi_++\psi_-}{\sqrt{2}},\;\;\;
    \phi_2=\frac{\psi_+-\psi_-}{\sqrt{2}},\;\;\;
    \bar{\phi}_1=\frac{\bar{\psi}_+-\bar{\psi}_-}{\sqrt{2}},\;\;\;
    \bar{\phi_2}=\frac{\bar{\psi}_++\bar{\psi}_-}{\sqrt{2}}.
\label{eq:rotation_K_fermions}    
\end{equation}
Applying this linear transformation of the fields $\psi_{\pm}$ to the Keldysh action in Eq.~\eqref{supeq:keldysh_pm} gives the action reported in Eqs.~(5) and (6) of the main text. The rotation also clarifies the role of the remaining Green's functions
\begin{subequations}
\begin{align}
    iG^R(x_1,x_2)&=\langle \phi_1(x_1)\bar{\phi}_1(x_2) \rangle=\Theta(t-t')\langle\{\,c(x_1), c^\dagger(x_2)\}\rangle,\label{eq:greens_rotated_1} \\
    iG^A(x_1,x_2)&=\langle \phi_2(x_1)\bar{\phi}_2(x_2) \rangle=-\Theta(t'-t)\langle\{\,c(x_1), c^\dagger(x_2)\}\rangle, \label{eq:greens_rotated_2} \\
    iG^K(x_1,x_2)&=\langle \phi_1(x_1)\bar{\phi}_2(x_2) \rangle= \braket{[c(x_1),c^{\dagger}(x_2)]}=-2\langle c^\dagger(x_2)c(x_1)\rangle +\langle\{\,c(x_1),c^\dagger(x_2)\}\rangle, \label{eq:greens_rotated_3}
\end{align}
\label{eq:greens_rotated}%
\end{subequations}
with $\braket{\phi_2(x_1)\bar{\phi}_1(x_2)}=0$ being zero as a consequence of trace preservation.
In fact, the retarded and the advanced Green's functions $G^R$ \eqref{eq:greens_rotated_1} and $G^A$ \eqref{eq:greens_rotated_2}, respectively, give information about the quasi-particle spectrum of the system, while the Keldsyh Green's function $G^K$ \eqref{eq:greens_rotated_3} gives information on the statistical occupation of the quasi-particle spectrum. The retarded and advanced Green functions are related by Hermitian adjoint $G^R(x_1,x_2)=[G^A(x_1,x_2)]^{\dagger}$ (complex conjugate and swap of the space-time variables $x_1$ and $x_2$), while $G^K$ is anti-hermitian $[G^K(x_1,x_2)]^{\dagger}=-G^{K}(x_1,x_2)$. The three Green functions \eqref{eq:greens_rotated} can be organised in the following 2x2 Keldysh-space matrix:
\begin{equation} 
\label{eq:propmat}
    \hat{G}(x_1,x_2) =
    \begin{pmatrix}
    G^R(x_1,x_2) & G^K (x_1,x_2)\\
    0 & G^A(x_1,x_2) 
    \end{pmatrix}.
\end{equation}
It is eventually customary to parametrise the Keldysh Green's function in terms of the hermitian distribution function $F(x_1,x_2)=[F(x_1,x_2)]^{\dagger}$, which is defined from the relation
\begin{equation}
\label{eq:distrib}
    G^K=G^R\circ F-F\circ G^A.
\end{equation}
In the previous equation, and in Sec.~\ref{sec:self_energy_calculations}, the symbol $\circ$ denotes the convolution product in the space-time variables. Whenever the symbol $\circ$ appears together with a propagator $\hat{G}$, as in Eq.~\eqref{eq:propmat}, contraction of the Keldysh 2x2 indices is also implied. The distribution function $F$ is the object of the kinetic Boltzmann equation, whose derivation we detail in the next section.  

\section{Self-energy and Boltzmann equation}
\label{sec:self_energy_calculations}

The evolution equation for the distribution function $F$ follows from the Dyson equation $[\hat{G}^{-1}_0-\hat{\Sigma}] \circ \hat{G}\,=\,\mathds{1}$. Here $\hat{G}_0$ denotes the bare propagators of the quadratic action $S_0$ in Eq.~(5) of the main text (we do not report their expressions as they are not needed in the derivation), while $\hat{G}$ the propagators dressed by the interaction via the self-energy $\hat{\Sigma}$. The latter has the same structure in Keldysh indices as the propagator $\hat{G}$ in Eq.~\eqref{eq:propmat} 
\begin{equation} 
\label{eq:sigmamat}
    \hat{\Sigma}(x_1,x_2) =
    \begin{pmatrix}
    \Sigma^R(x_1,x_2) & \Sigma^K (x_1,x_2)\\
    0 & \Sigma^A(x_1,x_2) 
    \end{pmatrix}.
\end{equation}
Taking the Keldysh component of the Dyson equation and parametrising $G^K$ in terms of $F$ as in Eq.~\eqref{eq:distrib}, one has
\begin{equation} 
\label{eq:kinetic}
    \left[-i(\partial_{t_1} +\partial_{t_2})-J(\nabla_{\vec{x_1}}^2-\nabla_{\vec{x_2}}^2)+(V(\vec{x_1})-V(\vec{x_2}))\right]F(x_1,x_2)=I[F]=\Sigma^K\circ\mathds{1}-(\,\Sigma^R\circ F-F\circ \Sigma^A\,).
\end{equation}
The latter is named kinetic equation, and the right-hand side $I[F]$ is called the collision integral. The self-energy matrix $\hat{\Sigma}$ encodes the effect of the interaction. Within this description, the computation of $F$ and of correlation functions necessarily amounts to calculating $\hat{\Sigma}$. The latter cannot be evaluated exactly and therefore one needs to develop an approximation scheme. In this manuscript, we consider the so-called Born perturbation theory approximation \cite{altland2010,sieberer2016,kamenev2011}. Within this approximation, $\hat{\Sigma}$ is evaluated by perturbatively expanding the interaction vertices around the quadratic part and thereby expressing the resulting Feynman diagrams in terms of the bare Green functions $\hat{G}_0$. 

As summarised in the main text, we consider, in particular, first-order contributions in $\Gamma$ to $\hat{\Sigma} \approx \hat{\Sigma}_{(1)}$. Such contributions are encoded into tadpole diagrams, where internal loops are connected to the external legs via a four-fielded vertex. There are four possible choices of the external legs, i.e., two possible fields for each of the two legs. Hence, we can sum up all internal contributions, and organise them in the 2x2 Keldysh-space matrix \eqref{eq:sigmamat}. Besides, at first-order in perturbation theory $\hat{\Sigma}(x_1,x_2) \approx \hat{\Sigma}_{(1)}(x_1,x_2)=\hat{\Sigma}_{(1)}(x)$ as tadpole graphs are defined at one space-time point $x_1=x_2=x$ only. Its components read:
\begin{equation}   \label{eq:self_en_ferm}
    \hat{\Sigma}_{(1)}(x)\,=\,
    \begin{pmatrix}
        \Sigma^R_{(1)}(x) & \Sigma^K_{(1)}(x) \\
        0 & \Sigma^A_{(1)}(x)
    \end{pmatrix}\,=\,\frac{\Gamma}{4}\,
    \begin{pmatrix}
        [f(G^A_0)+f(G^K_0)] & [2f(G^K_0)+f(G^A_0)-f(G^R_0)] \\
        0 & [f(G^R_0)-f(G^K_0)]
    \end{pmatrix}\,.
\end{equation}
Since the interaction vertices of the theory contain gradients of the fields $\phi_{1,2}$, the self-energy $\hat{\Sigma}$ is a differential operator acting on Feynman diagrams' external legs. In the previous equation, this is expressed by the appearance of the operator $f(G_0)$, which is defined as:
\begin{equation}
    f(G_0)=[\,G_0(x,x) \overleftarrow{\nabla}\,]\cdot\overrightarrow{\nabla} -[\,\overrightarrow{\nabla}G_0(x,x)\cdot\overleftarrow{\nabla}\,] + \overleftarrow{\nabla}\cdot [\,\overrightarrow{\nabla}G_0(x,x)\,] - \overleftarrow{\nabla}\cdot G_0(x,x) \overrightarrow{\nabla}.
\label{supeq:f_G_0}    
\end{equation}
In Eq.~\eqref{supeq:f_G_0}, the over-script arrows indicate the direction of the gradient operator. In particular, the notation $(G_0(x_1,x_2)\overleftarrow{\nabla})$ is used to denote differentiation of the Green's function $G_0(x_1,x_2)$ with respect to their second variable, i.e., $(G_0(x_1,x_2)\overleftarrow{\nabla})=(\vec{\nabla}_{x_2}G_0(x_1,x_2))$. Conversely, the notation $(\overrightarrow{\nabla}G_0(x_1,x_2))$ indicates differentiation with respect to the first variable of $G_0$, i.e., $(\overrightarrow{\nabla}G_0(x_1,x_2))=(\vec{\nabla}_{x_1}G_0(x_1,x_2))$. The differential operators in Eq.~\eqref{supeq:f_G_0} not enclosed into square brackets act on the diagrams' external legs, with $\overleftarrow{\nabla}$ differentiating with respect to the second variable of the external leg and $\overrightarrow{\nabla}$ with respect to the first one. From Eq.~\eqref{eq:self_en_ferm}, using the Hermitian adjoint properties of $G_0^{R,A,K}$ recalled after Eq.~\eqref{eq:greens_rotated}, it follows that $\Sigma^R_{(1)}=[\Sigma^A_{(1)}]^{\dagger}$ and $[\Sigma^{K}_{(1)}]^{\dagger}=-\Sigma^{K}_{(1)}$. The self-energy therefore shares the same structure as $\hat{G}$, as anticipated before Eq.~\eqref{eq:sigmamat} (this is expected to hold at any order in perturbation theory in $\Gamma$). We now derive a more useful expression for \eqref{eq:self_en_ferm} and \eqref{supeq:f_G_0}, which can be better used into the kinetic equation \eqref{eq:kinetic}. Since the self-energy matrix elements $\Sigma^K$, $\Sigma^R$ are convoluted with the identity and with $F$, respectively, we now want to turn operator $f(G_0)$ into a standard differential operator which acts on functions in subsequent positions on the right. We manage to do so by introducing a space-time variable $x_2$ which is constrained to take the same value of $x_1$ via a delta $\delta(x_1-x_2)$ (due to the tadpole structure of the diagrams). With this trick, we can write:
\begin{equation}
    \Sigma_{(1)}^R(x_1,x_2)=\frac{\Gamma}{4}\,\Big[\,\tilde{f}(G^A_0)+\tilde{f}(G^K_0)\,\Big], \quad \mbox{and} \quad     \Sigma_{(1)}^K(x_1,x_2)=\frac{\Gamma}{4}\,\Big[\,2\,\tilde{f}(G^K_0)+\tilde{f}(G^A_0)-\tilde{f}(G^R_0)\,\Big],
\label{supeq:sigma_1_R}    
\end{equation}
with
\begin{align}
    \tilde{f}(G_0)&=
    (\vec{\nabla}_{x_1}\delta(x_1-x_2)) \cdot [\,G_0(x_1,x_2) \vec{\nabla}_{x_2}-(\vec{\nabla}_{x_1} G_0(x_1,x_2))\,]+\nonumber \\
    &+\delta(x_1-x_2)\,[\,(\vec{\nabla}_{x_1} G_0(x_1,x_2))\cdot \vec{\nabla}_{x_2} + (\vec{\nabla}_{x_2}G_0(x_1,x_2))\cdot\vec{\nabla}_{x_2} 
    -(\nabla^2_{x_1} G_0(x_1,x_2))-(\vec{\nabla}_{x_1}\cdot\vec{\nabla}_{x_2} G_0(x_1,x_2))\,].
\end{align}
Derivatives of the Dirac deltas are understood in the sense of distribution, i.e., they have to be evaluated via integration by parts when inserted in the convolution products of the kinetic equation. We can also derive an analogous expression for $\Sigma^A$, which acts on the distribution function from the right, c.f. Eq.~\eqref{eq:kinetic}. 
where gradient operators act on the distribution function from the right only:
\begin{equation}
    \Sigma^A_{(1)}(x_1,x_2)=\frac{\Gamma}{4}\,\Big[\,\tilde{\tilde{f}}(G^R_0)-\tilde{\tilde{f}}(G^K_0)\,\Big],
\end{equation}
with
\begin{align}
    \tilde{\tilde{f}}(G_0)&=\delta(x_1-x_2)\,[\,(\nabla^2_{x_1} G_0(x_1,x_2))+(\nabla^2_{x_2} G_0(x_1,x_2))\,]\,-\,
    \overleftarrow{\nabla}_{x_1}\delta(x_1-x_2)\cdot[\,(\vec{\nabla}_{x_1} G_0(x_1,x_2))+(\vec{\nabla}_{x_2}G_0(x_1,x_2))\,]- \nonumber\\
    &+\,[\,(\vec{\nabla}_{x_2}G_0(x_1,x_2)) -\overleftarrow{\nabla}_{x_1} G_0(x_1,x_2)\,] \cdot (\vec{\nabla}_{x_2}\delta(x_1-x_2)).
\label{supeq:tilde_tilde_f}    
\end{align}
In order to extract the large-scale universal limit of the kinetic equation \eqref{eq:kinetic} and explicitly evaluate the convolution products appearing in the collision term, we resort to the Wigner transformation (WT). We first move to the set of Wigner coordinates $x=(x_1+x_2)/2$, $x'=x_1-x_2$, again using the 4-vector notation $x'=(\vec{x}',t')$ and $x=(\vec{x},t)$. These set of variables clearly distinguishes between the slow, macroscopic, evolution in $x$ and the fast dynamics in $x'$. The chain rule yields $\vec{\nabla}_{x_1}\,=\,\vec{\nabla}_x/2+\vec{\nabla}_{x'}$. The Wigner transform $A(x,k)$ corresponds to the Fourier transform of $A(x_1,x_2)$ with respect to the set of relative coordinates $x'$, which introduces the conjugated 4-momentum $k=(\vec{k},\epsilon)$:
\begin{equation}
    A(x,k)=\int d^dx'\,dt'\,A(x+x'/2,x-x'/2)\,e^{-i\vec{k}\cdot\vec{x}'+i\epsilon t'}, \, \, \, 
    A(x_1,x_2)=\int\,\frac{d^dk}{(2\pi)^d}\,\frac{d\epsilon}{2\pi}\,A\left(\frac{x_1+x_2}{2},k\right)\,e^{i\vec{k}\cdot(\vec{x_1}-\vec{x_2})-i\epsilon_k(t_1-t_2)}.
\label{supeq:WT_definition} 
\end{equation}
The main advantage of this procedure is to turn convolution products of two functions $A$ and $B$ in Eq.~\eqref{eq:kinetic} into Moyal products \cite{Moyal1,Moyal2,Moyal3,essler2022short,Moyal5} of their Wigner transforms
\begin{equation}
C(x_1,x_2)=(A \circ B)(x_1,x_2)=\int d^d x_3 A(x_1,x_3) B(x_3,x_2) \,\xlongrightarrow{\text{WT}}\,C(x,p)=A(x,k)\, \mathrm{exp}\left\{\frac{i}{2}[\overleftarrow{\partial}_x\overrightarrow{\partial}_k - \overleftarrow{\partial}_k\overrightarrow{\partial}_x]\right\}\,B(x,k).
\label{eq:Moyal_expansion}
\end{equation}

We now take the Euler-scaling limit \cite{spohn2012large,Dojon20}, where $\vec{x},t \to \infty$, $\Gamma \to 0$ with $\bar{\vec{x}}=\Gamma \vec{x}$ and $\bar{t}=\Gamma t$ fixed. In the reaction-limited regime $\hbar n \Gamma/J\ll 1$, time variations due to reactions are, indeed, slow on a scale $t \sim \Gamma^{-1}$. Space modulations, induced by trapping potential $V(\vec{x}\, \Gamma)$, are also slow on a macroscopic scale $\vec{x}\sim \Gamma^{-1}$. In this limit, the dependency of $G^{R,A,K}(x,x')$ on the macroscopic variable $x$ is much slower than the one on the relative one $x'$. In formulas, one has
\begin{align}
\vec{\nabla}_{x_1} G^{R,A,K}(x_1,x_2)=\frac{1}{2}\vec{\nabla}_{x}G^{R,A,K}(x,x')+\vec{\nabla}_{x'}G^{R,A,K}(x,x')&=\frac{\Gamma}{2}\vec{\nabla}_{\bar{x}}G^{R,A,K}(\bar{x},x')+\vec{\nabla}_{x'}G^{R,A,K}(\bar{x},x') \nonumber \\
&= \vec{\nabla}_{x'} G^{R,A,K}(\bar{x},x')\!+\!\mathcal{O}(\Gamma)\xlongrightarrow{WT}i\vec{k} G^{R,A,K}(\bar{x},k)\!+\!\mathcal{O}(\Gamma).
\label{eq:derivative_approximation}
\end{align}
Moreover, in the Euler-scaling limit, one can simplify the Moyal product in Eq.~\eqref{eq:Moyal_expansion} by expanding the complex exponential to first order in space-time derivatives. Derivatives of order $m>1$, indeed, lead to corrections $\mathcal{O}(\Gamma^m)$, which vanish in the Euler-scaling limit. 
From this expansion, the left-hand-side of the kinetic equation \eqref{eq:kinetic} becomes
\begin{equation} \label{eq:kinetic_term}
         \left[-i(\partial_{t_1} +\partial_{t_2})-J(\nabla_{\vec{x_1}}^2-\nabla_{\vec{x_2}}^2)+(V(\vec{x_1})-V(\vec{x_2}))\right]F(x_1,x_2) \xlongrightarrow{\text{WT}}\,i\Gamma \Big[
        \partial_{\bar{t}}+
        v_g(\vec{k})
        \cdot \vec{\nabla}_{\bar{x}}
        - \frac{1}{\hbar} \,\vec{\nabla}_{\bar{x}} V(\bar{\vec{x}})\cdot \vec{\nabla}_k\,\Big] F(\bar{x},k).
\end{equation}
We remark that Eq.~\eqref{eq:kinetic_term} is exact for harmonic potentials even away from the Euler-scaling limit. For anharmonic potentials, instead, there are corrections beyond the Euler-scaling limit. The first such correction to \eqref{eq:kinetic_term} is proportional to the cubic-space derivative of the potential \cite{Moyal5}. To deal with the WT of the collision integral, we first write $G_0^K=F(G_0^{R}-G_0^{A})$ in Eqs.~\eqref{supeq:sigma_1_R}-\eqref{supeq:tilde_tilde_f}. The previous equation follows from the WT of Eq.~\eqref{eq:distrib} at lowest order in derivatives. Here, albeit $G_0^{R,A,K}$ are bare Green functions, $F$ does not obey the free/quadratic-action dynamics, but it must be determined self-consistently from the kinetic equation. In the Euler-scaling limit, we can truncate the Moyal expansion \eqref{eq:Moyal_expansion} of convolution products at lowest order in derivatives and using Eq.~\eqref{eq:derivative_approximation}, we eventually obtain for the collision integral [$q=(\vec{q},\omega)$] 
\begin{equation}
    I[F]= 
    -\frac{i\Gamma}{2}\int\frac{d^dq}{(2\pi)^d}\frac{d\omega}{2\pi}(\vec{k}-\vec{q})^2[1-F(\bar{x},k)-F(\bar{x},q)+F(\bar{x},k)F(\bar{x},q)]A_0(\bar{x},q),
\label{supeq:collision_wigner}   
\end{equation}
where we introduced the bare spectral function $A_0(\bar{x},q)=-2\mbox{Im}(G^R_0(\bar{x},q))=i[G^R_0(\bar{x},q)-G^A_0(\bar{x},q)]$. One can also replace the bare green functions $G_0^{R,A}$ in \eqref{supeq:collision_wigner} with the dressed ones $G^{R,A}$, and therefore $A_0(\bar{x},q)\to A(\bar{x},q)$, within the self-consistent Born approximation. One then writes an equation for $G^{R,A}$ which has to be solved self-consistently together with the equation for $F$. The self-consistent Born approximation is necessary when the perturbative Born series of the self energy is divergent, as shown in Ref.~\onlinecite{buchhold2015kinetic}. In our case, however, the perturbative expression (of order $\Gamma$) for $\hat{\Sigma}_{(1)}$ in Eq.~\eqref{eq:self_en_ferm} is finite and the self-consistent Born approximation therefore only gives sub-leading corrections. For this reason, we do not consider it here.

From Eq.~\eqref{supeq:collision_wigner}, one can see that the collision integral $I[F]$ is a purely imaginary quantity: $F(\bar{y},k)$ is real since it is the WT of the hermitian distribution function $F(x_1,x_2)$. This result implies that the self-energy does not generate any renormalization of the quasi-particle energy spectrum $\epsilon_k(\bar{\vec{x}})=J\vec{k}^2+V(\bar{\vec{x}})$, i.e., the kinetic term in Eq.~\eqref{eq:kinetic_term} is left unchanged. On the contrary, the imaginary part of the collision integral represents the finite particle lifetime $\sim (n\Gamma)^{-1}$ introduced by the dissipative interactions. In the reaction-limited regime of weak dissipation, this lifetime $\sim (n\Gamma)^{-1}$ is much longer than the coherent time-scale set by $(J/\hbar)^{-1}$ and therefore quasi-particles remain well defined during time evolution. The spectral function $A(\bar{x},k) $ is accordingly sharply peaked in $\epsilon$ around $\epsilon_k(\bar{\vec{x}})$, with a width $\sim \Gamma$. In the non-interacting limit $\Gamma=0$, one exactly has $A_0(\bar{x},k)=2\pi \delta(\epsilon-\epsilon_{k}(\bar{\vec{x}}))$. 
This allows us to consider the on-shell distribution function $\widetilde{F}(\bar{\vec{x}},\vec{k},\bar{t})\equiv F(\bar{\vec{x}},\vec{k},\bar{t},\epsilon=\epsilon_{k}(\bar{\vec{x}}))$, which is given by 
\begin{equation}
    \widetilde{F}(\bar{\vec{x}},\vec{k},t)=\int\frac{d\epsilon}{2\pi}\,F(\bar{x},k)A(\bar{x},k)=i\int \frac{d\epsilon}{2\pi} G^K(\bar{x},k)= iG^K(\bar{\vec{x}},\vec{k},\bar{t},\bar{t})=1-2n(\bar{\vec{x}},\vec{k},\bar{t}).
\label{eq:F_on_shell}
\end{equation}
In the first equality, we used the peaked structure of the spectral function $A(\bar{x},k)$, in the second the expression of $G^K(\bar{x},k)$, in the third the inverse WT definition in $\epsilon$ \eqref{supeq:WT_definition}, and in the fourth equality the WT in space of Eq.~\eqref{eq:greens_rotated_3} (evaluated at equal times). The evolution equation for the one-body Wigner function $n(\bar{\vec{x}},\vec{k},\bar{t})$ is thus equivalent to that for $\widetilde{F}(\bar{\vec{x}},\vec{k},\bar{t})$. The latter is obtained by multiplying \eqref{eq:kinetic_term} and \eqref{supeq:collision_wigner} by $A(\bar{x},k)$ and integrating in $\epsilon$. The integral in $\omega$ in Eq.~\eqref{supeq:collision_wigner} is similarly handled by exploiting the Dirac delta form of $A_0(\bar{x},q)=2\pi \delta(\omega-\epsilon_k(\bar{\vec{x}}))$. Applying the substitution $\widetilde{F}(\bar{\vec{x}},\vec{k},\bar{t})=1-2 n(\bar{\vec{x}},\vec{k},\bar{t})$, in the Euler-scaling limit, the Boltzmann equation (8) of the main text eventually follows.

\section{Derivation of the asymptotic homogeneous decay in $d$ dimensions}
\label{sec:calculations_hom_decay}
The kinetic equation for the homogeneous fermionic particle density reads: 
\begin{equation} \label{eq:uuu}
    \partial_t n(\Vec{k},t)= -\Gamma \int \frac{d^dq}{(2\pi)^d}(\vec{k}-\vec{q})^2\,n(\Vec{k},t)\,n(\Vec{q},t)\,=-\Gamma \int \frac{d^dq}{(2\pi)^d}(k^2+q^2)\,n(\Vec{k},t)\,n(\Vec{q},t)\,.
\end{equation}
In the second equality, the term $2\,\Vec{k}\cdot\Vec{q}$ has been cancelled out as it vanishes in 
the integration over a spherically symmetric domain, whenever $n(\vec{k},t)=n(k,t)$ 
is also chosen to be spherically symmetric. 
This requires the initial distributions $n(\vec{k},0)=n(k,0)$ to be spherically symmetric. 
If this condition is satisfied, then spherical symmetry is preserved at any time $t>0$ 
and Eq.~\eqref{eq:uuu} applies. 
For the Fermi sea state considered in the main text, this is the case. Particles occupy all energy levels in a $d$-sphere of radius $k_d^F$ (the Fermi momentum). Because of the spherical symmetry of the initial state and of the collision integral, the $d$-dimensional problem can be recast into a one-dimensional radial problem. 
Hence, we define the following modified solid angle $\Theta_d$ and Fermi momentum $k_d^F$
\begin{equation}
    \Theta_d = \int d\Omega=\frac{d \sqrt{\pi}^d}{\Gamma(d/2+1)} \,, \quad k_d^F = 2\sqrt{\pi}[n_0 \Gamma(d/2+1)]^{1/d}.
\end{equation}
In the previous equations, $\Gamma(x)$ denotes the Euler-Gamma function of argument $x$ (not to be confused with the reaction constant $\Gamma$). We also introduce the following rescaled adimensional momenta:
\begin{equation}
    \Tilde{k}=\frac{k}{n_0^{1/d}} \,,\quad\quad 
    \Tilde{k}^F_d=\frac{k_d^F}{n_0^{1/d}} \,.
\end{equation}
$\tilde{k}_d^F$ is clearly a purely geometrical quantity and does not depend on the initial density. We can then write 
\begin{equation}
    n(t)= n_0 \frac{\Theta_d}{(2\pi)^d} \int_0^{\tilde{k}_d^F}d\Tilde{k}\, \Tilde{k}^{d-1}n(\Tilde{k},t)=
    n_0\, \Tilde{n}(t)\,,
\end{equation}
and define the rescaled density $\tilde{n}(t) = n(t)/n_0$.
Hence, the full dynamics is contained in the occupation function $n(\tilde{k},t)$ 
which is a function of the rescaled momenta and of the angular factor $\Theta_d$. 
Moreover, we can introduce the rescaled time $\tilde{t}=\Gamma n_0^{1+2/d} t$ so that we can then rewrite the collision integral in terms of adimensional variables:
\begin{equation}
    \partial_{\tilde{t}} n(\tilde{k}) = 
    -\frac{\Theta_d}{(2\pi)^d} \int_0^{\tilde{k}_d^F}
    d\Tilde{q}(\Tilde{q}^{d-1}\Tilde{k}^2+\Tilde{q}^{d+1})
    n(\Tilde{k})n(\Tilde{q})\,,
\end{equation}
where the temporal dependence $n (\tilde k,\tilde{t}) = n(\tilde k)$ has been left implicit for ease of notation. 
Following the method adopted in Ref.~\cite{Rosso2022} for the one-dimensional case, one 
obtains the following two equations:
\begin{equation}
\partial_{\tilde{t}}\tilde{n}(\tilde{t})=-2\frac{\Theta_d}{(2\pi)^d} \tilde{n}(\tilde{t}) \int_0^{\tilde{k}_d^F}
d\tilde{q}\tilde{q}^{d+1}n(\tilde{q},\tilde{t})\,, \qquad \frac{\partial_{\tilde{t}} n(\tilde{k},\tilde t)}{n(\tilde{k}, \tilde t)}= - \tilde{k}^2 \tilde{n}(\tilde{t}) +\frac{1}{2}\frac{\partial_{\tilde{t}}\tilde{n}(\tilde{t})}{\tilde{n}(\tilde{t})}\,.
\end{equation}
Integrating the latter equation leads to an exact expression for $n(\tilde{k},\tilde{t})$:
\begin{equation}
\label{eq:decay_MB}
    n(\tilde{k},\tilde{t})=\sqrt{\tilde{n}(\tilde{t})}\mathrm{exp}\Big[-\tilde{k}^2\int_0^{\tilde{t}}d\tilde{t}'\tilde{n}(\tilde{t}')\Big]\,.
\end{equation}
This expression is valid for momentum modes inside the Fermi sphere $\tilde{k}\leq\tilde{k}_d^F$, while outer modes do not populate. This can be seen from Eq.~\eqref{eq:uuu}, as the population $n(\tilde{k},\tilde{t})$ of a given mode is monotonically decreasing in time. 
Since $n(\tilde{k},\tilde{t})$ cannot be negative, if it is initially zero, then it remains zero at all times. 
Besides, Eq.~\eqref{eq:decay_MB} is fully determined only when the total density 
$\tilde{n}(\tilde{t})$ is known. However, the function 
\begin{equation} \label{eq:nu}
    \nu(\tilde{t})=\int_0^{\tilde{t}}d\tilde{t}'\tilde{n}(\tilde{t}')   
\end{equation}
is strictly increasing in the time variable, as the integrand function $\tilde{n}(\tilde{t}')$ is positive, entailing that Gaussian weights inside the Fermi sphere 
gain relevance as particles are lost. 
In order to determine the decay exponent, we can integrate Eq.~\eqref{eq:decay_MB} in the $\tilde{k}$ interval $[0,\tilde{k}_d^F]$. Dividing by $\sqrt{\tilde{n}(\tilde{t})}$ one arrives to:
\begin{equation} \label{eq:sqrtntilde}
    \sqrt{\tilde{n}(\tilde{t})}=
    \Theta_d I_{d-1}\Big(\nu(\tilde{t}), \tilde{k}_d^F\Big)\,,
\end{equation}
where we have defined the integral $I_p(a,b)=\int_0^{b}dx x^p e^{-ax^2}$. 
In the long-time limit,  the function $\nu(\tilde{t})$ diverges, entailing that we must evaluate $I_p(a,b)$ for positively diverging $a$. In this case, $I_p$ reduces to
\begin{equation}
    a\rightarrow\infty\Rightarrow I_p(a,b) \sim \alpha_{p+1}(p-1)!!(2a)^{-\frac{p+1}{2}}\,,
\end{equation}
with the definition $\alpha_{p+1}=1$ if $p+1$ is even, and $\alpha_{p+1}=\sqrt{\pi/2}$ if $p+1$ is odd.
Recalling Eq.~\eqref{eq:sqrtntilde} and the definition of $\nu(\tilde{t})$ such that 
$\tilde{n}(\tilde{t})=\partial_{\tilde{t}}\nu(\tilde{t})$, we find a differential equation for $\nu(\tilde{t})$ in terms of the integral $I_{d-1}(\nu, \tilde{k}_d^F)$, which greatly simplifies at asymptotically long times
and leads to:
\begin{equation}
    \nu(\tilde{t}) \sim \big\{ (d+1) [ \alpha_d\,(d-2)!! \Theta_d^2 (2\pi)^{-d}]^2 2^{-d} 
    \tilde{t} \big\}^{\frac{1}{d+1}}\,.
\end{equation}
The rescaled density is given by the time derivative of the last expression, namely:
\begin{equation}
    \tilde{n}(\tilde{t})\sim\Bigg\{\frac{[\alpha_d (d-2)!! \Theta_d]^2}
    {2^d (d+1)^d (2 \pi)^{2d}} \Bigg\}^{\frac{1}{d+1}}\,\tilde{t}^{-\frac{d}{d+1}}\,.
\end{equation}
The latter coincides with Eq.~(9) of the main text. Reintroducing the original variables one finally gets the long-time asymptotic of the particle density $n(t)$ in dimensionful units:
\begin{equation}\label{eq:ferm_homo}
    n(t) \sim n_0 \Bigg\{ \frac{1}{n_0^{d+2}} \frac{[ \alpha_d (d-2)!! \Theta_d]^2}
    {[(d+1)(8\pi^2) \Gamma ]^d} \Bigg\}^{\frac{1}{d+1}} t^{-\frac{d}{d+1}}\,.
\end{equation}
We note that the present derivation can be easily generalised to other initial states, e.g., thermal Gibbs states, identified by a spherically symmetric initial occupation function $n(k,0)$ by means of saddle-point analysis on Eq.~\eqref{eq:decay_MB}. In these cases, the same decay exponent $\tilde{t}^{-\frac{d}{d+1}}$ is found, while the form of $n(k,0)$ enters only in the amplitude of the decay.

\section{Effective exponent for the trap-release protocol}
\label{sec:effective_exp_supp}

\begin{figure}[t]
    \centering
    \includegraphics[width=0.5\textwidth]{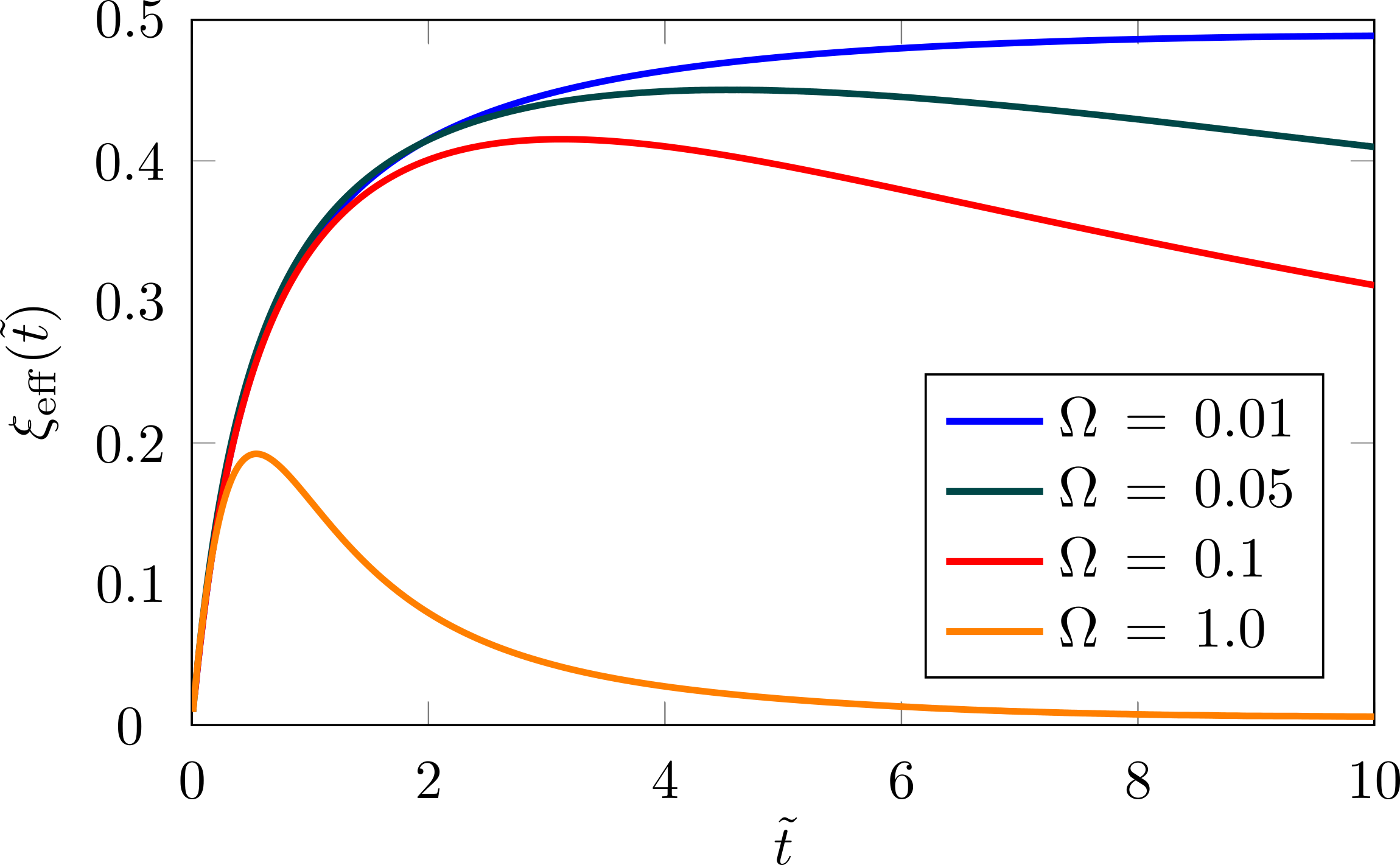}
    \caption{
    \textbf{Effective exponent for the trap-release quench}. Effective exponent $\xi_{\mathrm{eff}}$ for increasing values of $\Omega$ (from top to bottom), for the decay of the rescaled particle number $\tilde{N}(\tilde{t})$ as a function of the rescaled time interval $\tilde{t}\in[0,10]$. Particle number and time are rescaled as explained in the main text and the values of $\Omega$ reported are identical to those of Fig.~3(c) of the main text. The plot shows the existence of an approximate power-law decay $\tilde{N}(\tilde{t})\sim \tilde{t}^{-\xi_M}$ at intermediate times, where $\xi_{\mathrm{eff}}$ reaches a maximum $\xi_M$. For longer times, a non-algebraic decay sets in and $\xi_{\mathrm{eff}}$ decreases with respect to $\xi_M$. Such slow non power-law decay is present for every finite $\Omega$ value, and it takes place earlier in time as $\Omega$ is increased. All effective exponents have been computed with $b=1.1$ in Eq.~\eqref{supeq:effective_exp}. 
    }
    \label{fig:eff_exponent}
\end{figure}

We give here additional details concerning the decay of the particle number in the trap-release protocol discussed in Fig.~3(c)-(d) of the main text. As we presented in the main text, the behaviour of the re-scaled particle number $\tilde{N}(\tilde{t})=N(\tilde{t})/N_0$ at short rescaled times $\tilde{t}$ approaches the power-law $\tilde{N}(\tilde{t})\sim\tilde{t}^{-\xi}$, where $\xi$ is an exponent depending on parameter $\Omega$. In order to quantify $\xi$ and to see how the asymptotic decay changes as $\Omega$ is tuned, we compute the effective decay exponent $\xi_{\mathrm{eff}}$ \cite{hinrichsen2000non}
\begin{equation}
    \xi_{\mathrm{eff}}=-\,\frac{\log\,\Big[\tilde{N}(b\tilde{t})/\tilde{N}(\tilde{t})\Big]}{\log(b)},
    \label{supeq:effective_exp}
\end{equation}
with $b$ an arbitrary adimensional positive parameter. One can check that if $\tilde{N}(\tilde{t})$ asymptotically approaches a power law, i.e., $\tilde{N}(\tilde{t})\sim\tilde{t}^{-\xi}$ at long times, then $\lim_{t\rightarrow\infty}\xi_{\mathrm{eff}}=\xi$, $\forall b$.
In order to evaluate the effective decay exponent in a neighbourhood of a given time coordinate, we choose $b=1.1$. 

The plot in Fig.~\ref{fig:eff_exponent} shows the effective exponent $\xi_{\mathrm{exp}}(\tilde{t})$ in the interval $\tilde{t}\in[0,10]$ for various values of $\Omega$ (the same used in Fig.~3(c) of the main text). For very small $\Omega=0.01$ value, one can identify a power-law behaviour with effective exponent $\xi_{\mathrm{eff}}(\tilde{t})=\xi \approx 1/2$ throughout the considered domain, since $\xi_{\mathrm{eff}}$ converges to that value. Conversely, by increasing $\Omega$, an approximate algebraic decay $\tilde{N}(\tilde{t})\sim \tilde{t}^{-\xi_M}$ is observed, where the effective exponent reaches a maximum value $\xi_M$ and remains constant within an intermediate rescaled time window. For instance, the selected values $\Omega=0.05$, $\Omega=0.1$ are characterised by $\xi_M\approx 0.45$, $\xi_M\approx 0.4$ within the rescaled time interval $\tilde{t}\sim [3,6]$, $\tilde{t}\sim[2,4]$, respectively. For longer times the effective exponent $\xi_{\mathrm{eff}}$ drifts away from the aforementioned values and slowly decreases. This signals the onset of a non-algebraic (since $\xi_{\mathrm{eff}}$ is not constant) slow decay. As $\Omega$ is increased, the approximate decay exponent $\xi_{M}$ therefore decreases, the time window where it is observed is anticipated and shrinks. For $\Omega=0.01$, the non-algebraic behaviour takes place for longer times than those shown in the plot ($\tilde{t}<10$). Conversely, for large values $\Omega=1$, the intermediate power law decay is completely wiped out, no exponent can be identified and the non-algebraic decay immediately sets in. 

\end{document}